\documentclass[11pt,aps,showpacs,notitlepage]{revtex4-1}
\usepackage{amssymb,amsmath}
\usepackage{bm}
\usepackage[pdftex]{graphicx}
\usepackage{subfig}
\usepackage{color}
\usepackage{amsthm}
\usepackage[update,prepend]{epstopdf}

\newcommand{\beq}{\begin{equation}}
\newcommand{\beqn}{\begin{eqnarray}}
\newcommand{\eeq}{\end{equation}}
\newcommand{\eeqn}{\end{eqnarray}}

\begin{document}

\title{Preheating after multifield inflation with nonminimal couplings, II: \\ Resonance Structure}
\author{Matthew P. DeCross$^1$\footnote{Now at the Department of Physics and Astronomy, University of Pennsylvania.}, David I. Kaiser$^1$, Anirudh Prabhu$^1$\footnote{Now at the Department of Physics, Stanford University.}, Chanda Prescod-Weinstein$^2$, and Evangelos I. Sfakianakis$^3$\footnote{Now at NIKHEF and Leiden University.} }
\email{\baselineskip 11pt Email addresses: mdecross@sas.upenn.edu ; dikaiser@mit.edu ; aniprabhu@stanford.edu ; cprescod@uw.edu ; \\ evans@nikhef.nl}
\affiliation{$^1$Department of Physics, 
Massachusetts Institute of Technology, Cambridge, Massachusetts 02139 USA
\\
$^2${Department of Physics, University of Washington, Seattle, Washington 98195-1560}
\\
$^3$Department of Physics, University of Illinois at Urbana-Champaign, Urbana, Illinois 61801}
\date{\today}
\begin{abstract}
This is the second in a series of papers on preheating in inflationary models comprised of multiple scalar fields coupled nonminimally to gravity. In this paper, we work in the rigid-spacetime approximation and consider field trajectories within the single-field attractor, which is a generic feature of these models.
We construct the Floquet charts to find regions of parameter space in which particle production is efficient for both the adiabatic and isocurvature modes, and analyze the 
 resonance structure using analytic and semi-analytic techniques.
Particle production in the adiabatic direction is characterized by the existence of an asymptotic scaling solution at large values of the nonminimal couplings, $\xi_I \gg 1$, in which the dominant instability band arises in the long-wavelength limit, for comoving wavenumbers $k \rightarrow 0$.
However, the large-$\xi_I$ regime is not reached until $\xi_I \geq {\cal O} (100)$. In the intermediate regime, with $\xi_I \sim {\cal O}(1 - 10)$, the resonance structure depends strongly on wavenumber and couplings. 
The resonance structure for isocurvature perturbations is distinct and more complicated than its adiabatic counterpart. An intermediate regime, for $\xi_I \sim {\cal O} (1 - 10)$, is again evident. 
For large values of $\xi_I$, the Floquet chart consists of densely spaced, nearly parallel instability bands, suggesting a very efficient preheating behavior. The increased efficiency arises from features of the nontrivial field-space manifold in the Einstein frame, which itself arises from the fields' nonminimal couplings in the Jordan frame, and has no analogue in models with minimal couplings.
Quantitatively, the approach to the large-$\xi_I$ asymptotic solution for isocurvature modes is slower than in the case of the adiabatic modes. 
\end{abstract}
\pacs{98.80.Cq ; 95.30.Cq.  Preprint MIT-CTP/4848.}
\maketitle

\section{Introduction} 

This paper is the second of a three-part series focusing on the preheating dynamics of multifield models of inflation with nonminimal couplings. (See also Refs.~\cite{MultiPreheat1,MultiPreheat3}.) Such models are well-motivated: realistic models of high-energy physics generically include many scalar degrees of freedom at high energy scales \cite{LythRiotto,WandsReview,MazumdarRocher,MartinRingeval,VenninWands,GongMultifield}, and nonminimal couplings are generically required as renormalization counterterms for interacting scalar fields in curved spacetime \cite{Callan,Bunch,BirrellDavies,Buchbinder,ParkerToms,Odintsov1991,Markkanen2013}. The family of models we consider includes well-known examples like Higgs inflation \cite{BezrukovShaposhnikov} (see also \cite{Futamase,Salopek,Fakir,Makino,DKconstraints,DKnGET}), as well as related models with attractor-like solutions \cite{KMS,GKS,KS,SSK,Lindealpha}.

The epoch of post-inflation reheating is critical for several reasons. (For reviews of reheating, see Refs.~\cite{GuthKaiser,BTW,AllahverdiRev,Frolov,AHKK,MustafaBaumann}.) Reheating is responsible for populating the universe with Standard Model particles in thermal equilibrium. Moreover, understanding the dynamics of reheating is essential for connecting predictions from inflationary models with high-precision measurements of primordial perturbation spectra, since reheating affects the expansion history of the universe between the end of inflation and eras such as big-bang nucleosynthesis~\cite{AdsheadEasther,Dai,Creminelli,MartinReheat,GongLeungPi,CaiGuoWang,Cook,Heisig}. 

The first stage of reheating is characterized by the resonant decay of the scalar-field condensate(s) that had driven inflation. We use the doubly-covariant formalism introduced in Ref.~\cite{MultiPreheat1}, which self-consistently incorporates metric perturbations and field fluctuations to first order and maintains reparameterization freedom of the nontrivial field-space manifold. In this paper we focus on the early stage of preheating, working to linear order in the fluctuations. To simplify, we only consider the decay of the inflaton into non-gauged scalar fields, for now ignoring higher-spin particles such as fermions and gauge fields, although we expect that such fields will have a rich phenomenology \cite{AHKK,KLS,GreeneKofmanfermions,Peloso,TsujikawaBassettViniegra,MarotoMazumdar,KKLvP,Davis,GarciaBellido,HiggsReheat1,HiggsReheat2,HiggsReheat3,AminOscillons,Allahverdi,HiggsReheat4,Deskins,MPHBaryogenesis,Mustafa,HertzbergKarouby,Adshead:2015kza,Adshead:2015pva,Adshead:2015jza,McDonough,MustafaWires,Lozanov,ZhouOscillons,MustafaDuration,Figueroa}. Our study builds upon previous studies of preheating in models with nonminimal couplings and/or noncanonical kinetic terms~\cite{BassettGeometric,ShinjiPapers,ShinjiBassett,WatanabeWhite,NonCanKin1,NonCanKin2,NonCanKin3,NonCanKin4,Racioppi,VanDeBruck,Ema}, with the aim of identifying characteristic features across a wide range of parameters.

In Ref.~\cite{MultiPreheat1}, we identified several distinctions between preheating in multifield models with nonminimal couplings compared to more familiar, minimally coupled models. First, the oscillations of the background fields are affected by the conformal stretching of the fields' potential in the Einstein frame. Second, the single-field attractor behavior tends to enhance efficiency during preheating compared to multifield, minimally coupled models \cite{Barnaby,Battefeld}. Third, in the limit of strong couplings, $\xi_I \gg 1$, the nontrivial field-space manifold affects the effective masses for adiabatic and isocurvature fluctuations differently, which in turn affects the energy transfer to long-wavelength perturbations. In the present paper we analyze the structure of resonances in this family of models semi-analytically and numerically across wide regions of parameter space. 

In this paper we adopt the approximation of a rigid spacetime, in which we imagine keeping the energy density fixed while sending $M_{\rm pl} \rightarrow \infty$. (The reduced Planck mass is given by $M_{\rm pl} \equiv 1/\sqrt{8 \pi G} = 2.43 \times 10^{18}$ GeV.) In that limit, we may neglect the expansion of spacetime during preheating, as well as the effects of the coupled metric perturbations \cite{AHKK}. Then the oscillations of the inflaton condensate(s) become periodic, and we may apply the tools of Floquet theory to study the resonant amplification of perturbations. In Section \ref{BackgroundDynamics} we examine the background dynamics for a two-field model after inflation, highlighting distinctions between oscillations during preheating with and without nonminimal couplings. We analyze the spectral content of the background oscillations; as shown in Section \ref{BackgroundDynamics}, analytic progress can be made in the limit of large nonminimal couplings. 

Knowledge of the behavior of the background fields during the oscillating phase is critical for understanding the resonant production of particles during preheating.
 In Section \ref{EvolutionFluctuations} we briefly review the covariant mode expansion for the fluctuations introduced in Ref.~\cite{MultiPreheat1}.
  Using this covariant formalism, we may construct Floquet charts and identify the instability bands for the fluctuations.
In Section \ref{Adiabaticmodes} we analyze the behavior of adiabatic modes and find a universal scaling form for the Floquet chart which emerges in the limit of large nonminimal couplings. 
In Section \ref{Isocurvaturemodes} the more complex case of isocurvature modes is presented. We construct the three-dimensional Floquet charts for a wide range of nonminimal couplings and potential parameters. In the region of large nonminimal couplings, the Floquet chart is comprised of a dense set of nearly parallel instability bands, indicating a very efficient amplification of isocurvature modes. A scaling solution for large $\xi_I$ is found, similar to the adiabatic case, although the asymptotic limit is reached more slowly for comparable values of $\xi_I$. 
Concluding remarks follow in Section \ref{Conclusions}.
  
\section{Background Dynamics}
\label{BackgroundDynamics}

Following the analysis of Refs. \cite{MultiPreheat1,KMS,GKS,KS,SSK} we study inflationary models with multiple scalar fields coupled nonminimally to the spacetime Ricci scalar. The formalism developed in those studies may be applied to arbitrary numbers of fields, and in previous work \cite{KS} we confirmed that the single-field attractor behavior in these models holds for cases with more than two fields. Nonetheless, in this paper we restrict attention to models with just two fields, $\phi$ and $\chi$, which significantly simplifies visualizing various results. We work in $3+1$ spacetime dimensions and choose the mostly plus spacetime metric $(-,+,+,+)$. 

We begin with the action in the Jordan frame, in which the fields' nonminimal couplings are explicit:
\beq
S = \int d^4 x \sqrt{-\tilde{g} } \left[ f (\phi^I ) \tilde{R} - \frac{1}{2} \delta_{IJ} \tilde{g}^{\mu\nu} \partial_\mu \phi^I \partial_\nu \phi^J - \tilde{V} (\phi^I ) \right] .
\label{SJ}
\eeq
We use upper-case Latin letters to label field-space indices, $I, J = 1,2$. For the remainder of this paper we will consider a two-field model with $\phi^I = \{ \phi , \chi \}^T$. Greek letters label spacetime indices, $\mu, \nu = 0, 1, 2, 3$, and tildes denote Jordan-frame quantities. 
By performing the conformal transformation 
\beq
 \tilde{g}_{\mu\nu} (x) \rightarrow g_{\mu\nu} = \frac{2}{M_{\rm pl}^2} f \left( \phi^I (x) \right) \> \tilde{g}_{\mu\nu} (x),
\label{gtildeg}
\eeq
we can bring the gravitational part of the action to the usual Einstein-Hilbert form \cite{DKconf,Abedi}
\beq
S = \int d^4 x \sqrt{-g} \left[ \frac{ M_{\rm pl}^2}{2} R - \frac{1}{2} {\cal G}_{IJ} (\phi^K ) g^{\mu\nu} \partial_\mu \phi^I \partial_\nu \phi^J - V (\phi^I ) \right] .
\label{SE}
\eeq
The induced field-space metric is given by
\beq
{\cal G}_{IJ} (\phi^K ) = \frac{ M_{\rm pl}^2 }{2 f (\phi^K) } \left[ \delta_{IJ} + \frac{3}{ f (\phi^K) } f_{, I} f_{, J} \right] ,
\label{GIJ}
\eeq
where $f_{, I} = \partial f / \partial \phi^I$. Explicit components of ${\cal G}_{IJ} (\phi^K)$ for our two-field model may be found in Appendix A of Ref.~\cite{MultiPreheat1}. We note that models that have canonical kinetic terms in the Jordan frame will nonetheless develop a nontrivial field-space manifold in the Einstein frame \cite{DKconf}.
The potential in the Einstein frame is similarly stretched by a conformal factor,
\beq
V (\phi^I) = { M_{\rm pl}^4 \over 4 f^2 (\phi^I) } \tilde{V} (\phi^I ) .
\label{VE}
\eeq
Renormalization of models with self-coupled scalar fields in curved spacetime requires counter-terms of the form $\xi \phi^2 R$ for each nonminimally coupled field \cite{Callan,Bunch,BirrellDavies,Buchbinder,ParkerToms,Odintsov1991,Markkanen2013}. We therefore take $f (\phi^I)$ to be of the form
\beq
f (\phi, \chi ) = \frac{1}{2} \left[ M_{\rm pl}^2 + \xi_\phi \phi^2 + \xi_\chi \chi^2 \right] .
\label{f2field}
\eeq
Each scalar field $\phi^I$ couples to the Ricci scalar with its own nonminimal-coupling constant, $\xi_I$, which we take to be positive. 

The form of the potential can be arbitrary. We adopt a simple polynomial potential in the Jordan frame that includes the highest renormalizable interaction terms for scalar fields in $3 + 1$ spacetime dimensions:
\beq
\tilde{V} (\phi, \chi) = \frac{\lambda_\phi}{4} \phi^4 + \frac{g}{2}  \phi^2 \chi^2 + \frac{ \lambda_\chi}{4} \chi^4 .
\label{VJphichi}
\eeq
In order for the potential to be non-tachyonic, we must choose $\lambda_I >0$. Furthermore we neglect bare masses $m_I^2$, the effects of which can be studied using the methods developed here and in \cite{MultiPreheat1}. On a more practical level, keeping only the quartic potential allows us to compare the resonances and Floquet charts of nonminimally coupled models to their well-studied minimally coupled counterparts~\cite{Boyanovsky,KaiserPreh2,GKLS}, and identify features that arise due to the nontrivial field-space metric, which is a manifestation of the nonminimal couplings in the Einstein frame.

As discussed in Section II B of Ref.~\cite{SSK}, observational constraints place restrictions on combinations of couplings at the high energy scales of inflation. In particular, in models like Higgs inflation \cite{BezrukovShaposhnikov}, one typically finds $\lambda_I \sim {\cal O} ( 10^{-2} - 10^{-4} )$ at the energy scales of inflation (the range stemming from uncertainty in the value of the top-quark mass, which affects the running of $\lambda_I$ under renormalization-group flow) \cite{Barvinsky,BezrukovRunning,Allison,MossXi}. The range of $\lambda_I$, in turn, requires $\xi_I \sim {\cal O} ( 10^2 - 10^4 )$ at high energies, in order to fit various observational constraints from primordial curvature perturbations --- a reasonable range, given that $\xi_I$ typically rises with energy scale under renormalization-group flow, with no UV fixed point \cite{Buchbinder}. Even with such large values of $\xi_I$, the inflationary dynamics occur at energy scales well below any nontrivial unitarity cut-off scale. (See Ref.~\cite{BezrukovRunning} and references therein for further discussion.)

We are interested in the structure of resonances during the preheating phase for this family of models, and hence we consider a broader range of values for the couplings, especially $\xi_I$. Many features of the resonance structure vary markedly with $\xi_I$. In particular, we explore three regimes of interest: $0 < \xi_I \leq {\cal O} (1)$, $\xi_I \sim {\cal O} (1 - 10)$, and $\xi_I \geq {\cal O} (100)$. We also analyze the asymptotic behavior in the strongly coupled limit, $\xi_I \gg 1$.

\subsection{Potential Geometry and Single-Field Attractor}
\label{AttractorBehavior}

\begin{figure}
\centering
\includegraphics[width=0.48\textwidth]{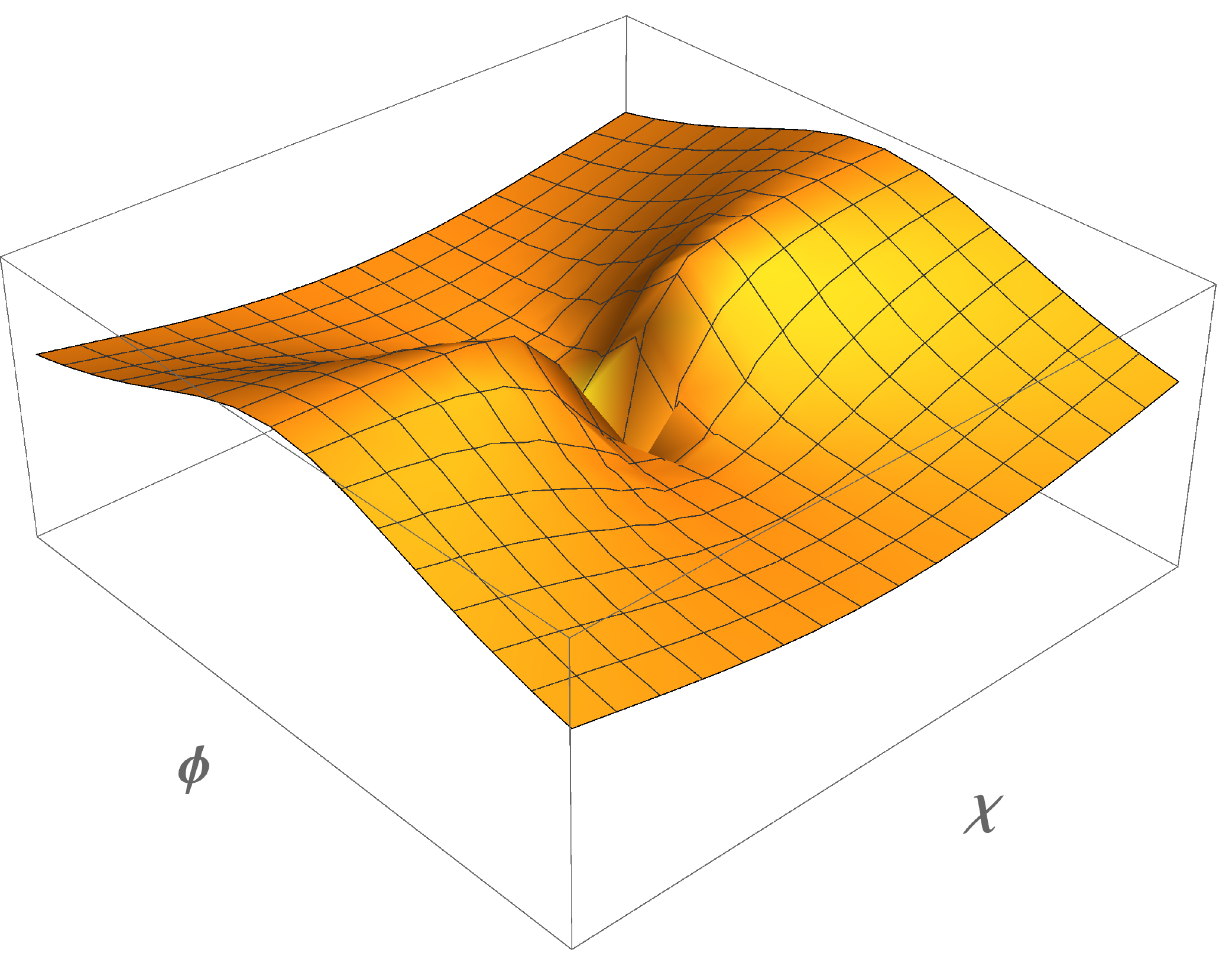} \quad \includegraphics[width=0.48\textwidth]{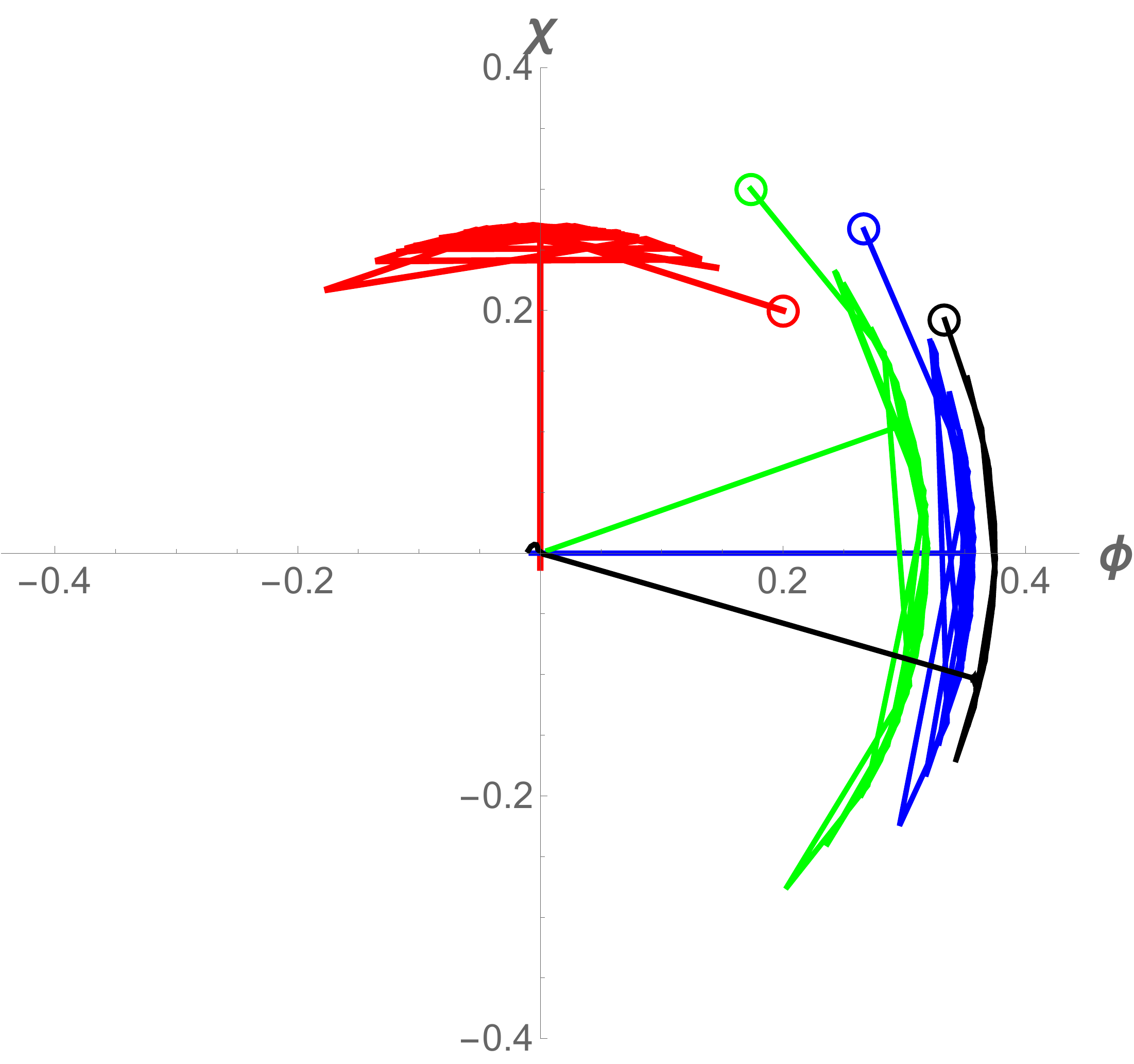}
\caption{ \small \baselineskip 12pt ({\it Left}) Potential in the Einstein frame, $V (\phi^I)$, for a two-field model with $\lambda_\chi = 1.25 \>\lambda_\phi $, $g = \lambda_\phi$, and $\xi_\chi = 0.8\> \xi_\phi$. ({\it Right}) Field trajectories for different couplings and initial conditions (from Ref.~\cite{KS}). Open circles indicate fields' initial values (in units of $M_{\rm pl})$. We set the fields' initial velocities to zero and adjust the initial angle in field space, $\theta_0 = {\rm arctan} (\phi_0 / \chi_0)$. We fix $\lambda_\phi = 10^{-2}$ and $\xi_\phi = 10^3$ and vary the other parameters $\{ \lambda_\chi , g, \xi_\chi, \theta_0\}$ as follows: $\{ 0.75  \lambda_\phi , \lambda_\phi, 1.2 \xi_\phi , \pi / 4 \}$ (red), $\{ \lambda_\phi, \lambda_\phi , 0.8 \xi_\phi , \pi / 4 \}$ (blue), $\{\lambda_\phi, 0.75  \lambda_\phi , 0.8 \xi_\phi , \pi / 6 \}$ (green), and $\{ \lambda_\phi , 0.75 \lambda_\phi , 0.8 \xi_\phi , \pi / 3 \}$ (black). In each case, the initial transient motion damps out within a few efolds, yielding effectively single-field evolution during inflation.}
\label{VEfig}
\end{figure}

Inflation occurs in a regime in which $\xi_J (\phi^J )^2 \gg M_{\rm pl}^2$ for at least one component, $J$ (for $\xi_J \ll 1$ the condition becomes $\phi^J \gg  M_{\rm pl}$). The potential in the Einstein frame becomes asymptotically flat along each direction of field space, as each field $\phi^I$ becomes arbitrarily large:
\beq
V (\phi^I) \rightarrow \frac{M_{\rm pl}^4}{4} \frac{\lambda_I}{\xi_I^2} \left[ 1 + {\cal O} \left( \frac{ M_{\rm pl}^2}{\xi_I (\phi^I)^2 } \right) \right]
\label{Vasympt}
\eeq
(no sum on $I$). Unless some explicit symmetry constrains all coupling constants in the model to be identical ($\lambda_I = g = \lambda$, $\xi_I = \xi$), then the potential in the Einstein frame will develop ridges and valleys, as shown in Fig.\ \ref{VEfig}.
We restrict the parameter space to $-g < \sqrt{\lambda_\phi \lambda_\chi}$, where the potential in both frames (Jordan and Einstein) is positive at all times. Since both the ridges and the valleys satisfy $V > 0$, the universe will inflate (albeit at different rates) whether the fields evolve along a ridge or a valley toward the global minimum of the potential. 

As the right side of Fig.\ \ref{VEfig} makes clear, multifield models with nonminimal couplings display strong single-field attractor behavior during inflation, across a wide range of couplings and initial conditions \cite{MultiPreheat1,KS,SSK}. If the fields happen to begin evolving along the top of a ridge, they will eventually fall into a neighboring valley at a rate that depends on the local curvature of the potential \cite{KMS, SSK}. Once the fields fall into a valley, Hubble drag quickly damps out any transverse motions in field space within a few efolds, after which the system evolves with very little turning in field space for the remainder of inflation \cite{KMS,GKS,KS,SSK}. 

The orientation of the valley in field space, $\theta = {\rm arctan} (\phi / \chi)$, depends on the choice of couplings \cite{KMS}.
As demonstrated in Ref.~\cite{MultiPreheat1}, we may always exploit the covariant framework and perform a rotation in field space, $\phi^I \rightarrow \phi^{I \prime}$, such that the valley lies along the direction $\chi^\prime = 0$ during inflation. Once the fields settle into the single-field attractor, the field-space metric becomes effectively diagonal: ${\cal G}_{\phi^\prime \chi^\prime} = {\cal O} (\chi^{\prime }) \sim 0$.

The fields will only remain near the top of a ridge for a significant duration of inflation if {\it both} the local curvature of the potential {\it and} the initial conditions in field-space $\chi (t_0)$ are tuned to be exponentially close to zero. In those cases, the system's evolution during the last 60 efolds of inflation can amplify non-Gaussianities and isocurvature perturbations, which could potentially be observable \cite{KMS,SSK,SebastianRP}. However, if neither the local curvature of the potential nor the fields' initial conditions are exponentially fine-tuned, then the system will rapidly relax into a valley and evolve along an effectively single-field trajectory right to the end of inflation. 
Within the single-field attractor, these models predict values for spectral observables such as the primordial spectral index and its running ($n_s$ and $\alpha$), the ratio of power in tensor to scalar modes ($r$), primordial non-Gaussianity ($f_{\rm NL}$), and the fraction of power in isocurvature rather than adiabatic scalar modes ($\beta_{\rm iso}$) all in excellent agreement with the latest observations \cite{KMS,GKS,KS,SSK}.

We have extensively studied the geometry of the potential and its parameter dependence in Ref.~\cite{MultiPreheat1}. As discussed there, the single-field attractor behavior in these models persists after the end of inflation and into the early phase of preheating, at least to linear order in the fluctuations. For the remainder of this work, we will therefore only consider scenarios in which the background fields evolve within a single-field attractor. Without loss of generality, we consider that attractor to lie along the $\chi=0$ direction in field space.

\subsection{Spectral Content of the Oscillating Background Field}
\label{BackgroundSpectral}

Given the dependence of the effective equation of state on $\xi_\phi$ while the background field(s) oscillate \cite{MultiPreheat1}, we expect the oscillations themselves to show significant deviation from the case of minimal couplings. We demonstrated in Ref.~\cite{MultiPreheat1} that the frequency of oscillation $\omega$ exceeds the Hubble expansion rate during preheating in these models, $\omega / H > 1$, across the entire range $10^{-3} \leq \xi_\phi \leq 10^3$ (see Fig. 10 in Ref.~\cite{MultiPreheat1}). To facilitate comparison with the well-studied case of a minimally coupled field with quartic self-coupling \cite{Boyanovsky,KaiserPreh2,GKLS}, 
we therefore neglect Hubble expansion during the oscillating phase (though its effects may be incorporated perturbatively \cite{KaiserPreh1}). We further neglect backreaction from produced particles on the background fields' oscillation. Hence we may employ Fourier analysis to study the dependence of the harmonic structure of the background oscillations on $\xi_\phi$.

Although we are working in the rigid-spacetime approximation, we aim to be able to exploit as many results as possible when considering the more realistic case of an expanding universe in Ref.~\cite{MultiPreheat3}. Hence it is important to examine the range of background field values that are relevant. Fig.\ \ref{fig:phibackground} shows the behavior of the inflaton background field $\phi (t)$ after the end of inflation, neglecting backreaction. For $\xi_\phi \gtrsim 1$ inflation ends at $\phi \approx 0.8 M_{\rm pl} /\sqrt{\xi_\phi}$  \cite{MultiPreheat1}. However, using the rigid-spacetime approximation to study oscillations with amplitude $\phi_{\rm max} = 0.8 M_{\rm pl} /\sqrt{\xi_\phi}$ introduces a considerable error, since at the end of inflation the Hubble friction term is still non-negligible, leading to a decrease of the inflaton amplitude by a factor of $2$ within one oscillation. The most relevant regime for calculating quantities like the period and spectral content of the oscillating inflaton field is therefore $\phi_{\rm max} = (\alpha \times 0.8) M_{\rm pl} / \sqrt{\xi_\phi}$ with $0.3 \lesssim \alpha \lesssim 0.5$. These values will capture particle production during the first few oscillations of the inflaton field, for which we may safely neglect backreaction effects. We will restrict our analysis for the remainder of this work to $\phi_{\rm max} = ( 0.5 \times 0.8 ) M_{\rm pl} / \sqrt{\xi_\phi}$, unless stated otherwise.

\begin{figure}
\centering
\includegraphics[width=.48\textwidth]{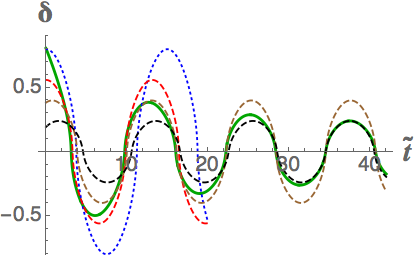} \quad \includegraphics[width=.48\textwidth]{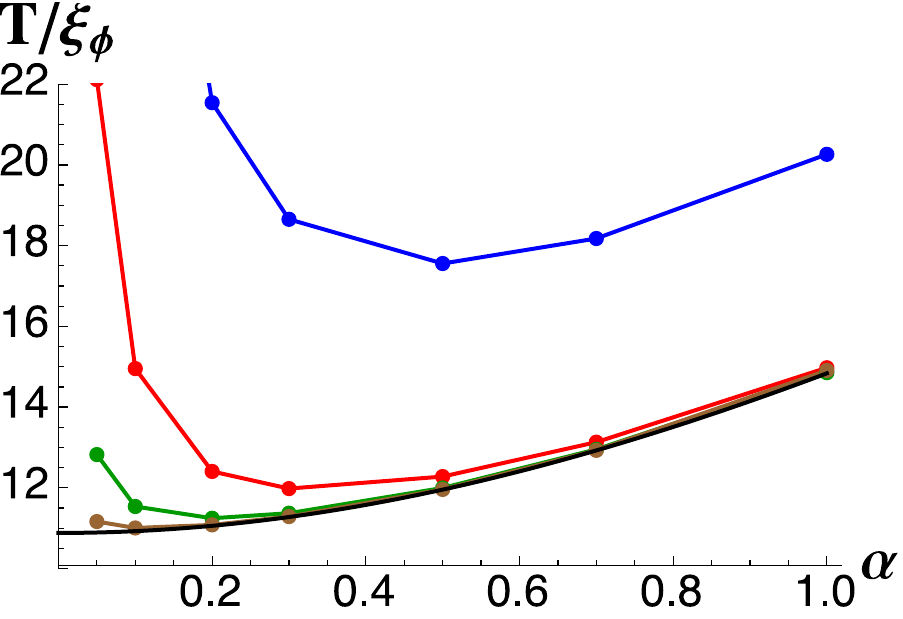}
\label{fig:phibackground}
\caption{\small \baselineskip 11pt ({\it Left}) The rescaled background solution $\delta(t) = \sqrt{\xi_\phi}\,\phi(t) / M_{\rm pl}$ as a function of $\tilde{t} = \sqrt{\lambda_\phi} \, M_{\rm pl} \, t / \xi_\phi$ after the end of inflation, with $\xi_\phi = 100$. Shown in green is the evolution of $\delta (t)$ when Hubble expansion is included self-consistently. The other curves show the rigid-spacetime solution with initial condition $\delta(0) =(\alpha \times 0.8) \sqrt{\xi_\phi} \> \phi_0 / M_{\rm pl}$, with $\alpha = 1,0.7, 0.5, 0.3$ (blue, red, brown, and black, respectively).
({\it Right}) The rescaled period of background oscillation in a static universe (in units of $(\sqrt{\lambda_\phi} \, M_{\rm pl})^{-1}$) for $\xi_\phi=10,10^2,10^3,10^4$ (blue, red, green, and brown, respectively) as a function of the amplitude parameter $\alpha$. The analytic approximation for the period in the large-$\xi_\phi$ regime is shown in black, and is derived under the assumption that $6 \xi_\phi \alpha^2 \gg 1$.  }
\label{fig:phibackground}
\end{figure}

Within the single-field attractor, with $H \sim 0$, the background field $\phi (t)$ obeys the equation of motion
\beq
\ddot{\phi} + \Gamma^\phi_{\> \phi \phi} \dot{\phi}^2 + {\cal G}^{\phi \phi} V_{, \phi} \simeq 0 .
\label{phieomnoH}
\eeq
We rescale $\tau \equiv \sqrt{\lambda_\phi} \> t$, so that the dynamics depend only on $\xi_\phi$. After inflation ends at $\tau_{\rm end}$, $\phi (\tau)$ oscillates with a period given by
\beq
T  = 2 \int_{- \phi_0}^{\phi_0} d \phi \sqrt{ \frac{ {\cal G}_{\phi \phi} }{2 V (\phi_0) - 2 V (\phi) }  }.
\label{T}
\eeq
(We label $\phi_0 = \phi (\tau_{\rm end})$ as the amplitude of the field at the start of preheating.) 
The period behaves as $T \propto \xi_\phi$ for $\xi_\phi \gg1 $, as calculated in Appendix B of Ref.~\cite{MultiPreheat1} and shown in Fig.\ \ref{fig:phibackground}. In the limit $\xi_\phi \gg 1$, the period $T$ rises monotonically with $\alpha$, though for intermediate values of $\xi_\phi$ the period shows a more complicated dependence on $\alpha$.

\begin{figure}
\centering
\includegraphics[width=0.48\textwidth]{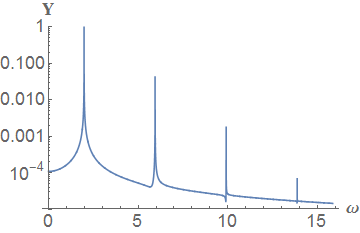} \includegraphics[width=0.48\textwidth]{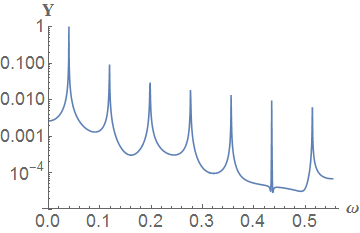}
\caption{ \small \baselineskip 12pt  Fast Fourier Transforms of $\phi (\tau)$ for $\xi_\phi = 0$ ({\it left}) and $\xi_\phi = 10$ ({\it right}). For $\xi_\phi > 0$, the background field's oscillations show a richer spectral content, including more non-negligible harmonics, which may drive additional resonances compared to the $\xi_\phi = 0$ case. All higher harmonics are normalized with respect to the amplitude of the fundamental mode and $\omega$ is measured in units of $\sqrt{\lambda_\phi} \, M_{\rm pl}$.}
\label{phiFourierfig}
\end{figure}

The terms in Eq. (\ref{phieomnoH}) that arise from the nontrivial field-space metric affect the harmonic structure of $\phi$'s oscillations. In the limit $\xi_\phi = 0$, with the Jordan-frame potential of Eq. (\ref{VJphichi}), Eq. (\ref{phieomnoH}) may be solved analytically as a Jacobian elliptic cosine \cite{Boyanovsky,KaiserPreh2,GKLS}: $\phi (t) = \phi_0 \> {\rm cn} (\phi_0 \tau , 1 / \sqrt{2})$. The function ${\rm cn} \> (x, \kappa)$ is periodic with period $4 K (\kappa)$, where $K (\kappa)$ is the complete elliptic integral of the first kind \cite{AbramSteg}. The Jacobian elliptic cosine may be expanded in the infinite series (see Eq. 16.23.2 of \cite{AbramSteg})
\beq
{\rm cn} \> (x , \kappa) = \frac{ 2 \pi}{\sqrt{\kappa} \> K (\kappa ) } \sum_{n = 0}^\infty \frac{ q^{n + 1/2} }{1 + q^{2n + 1} } \cos (4 n + 2) v ,
\label{cncos}
\eeq
where $q \equiv \exp [ - \pi K^\prime / K]$, $v \equiv \pi x / (4 K)$, and $K^\prime (\kappa) = K (1 - \kappa)$. Given $\kappa = 1/ \sqrt{2}$ for $\xi_\phi = 0$, we find $q = 0.076$, and hence terms with $n \geq 3$ in the series expansion enter with coefficients ${\cal O} (10^{-3})$ or less. In the case of minimal coupling, in other words, the oscillations of $\phi ( \tau )$ are well approximated by a dominant cosine term and a first harmonic. As shown in Fig.\ \ref{phiFourierfig}, for $\xi_\phi > 0$ the harmonic structure shifts, with (in general) more non-negligible harmonics. This can lead to regions of enhanced resonance during preheating, as we will see in Sections \ref{Adiabaticmodes} and \ref{Isocurvaturemodes}.

\begin{figure}
\centering
\includegraphics[width=0.65 \textwidth]{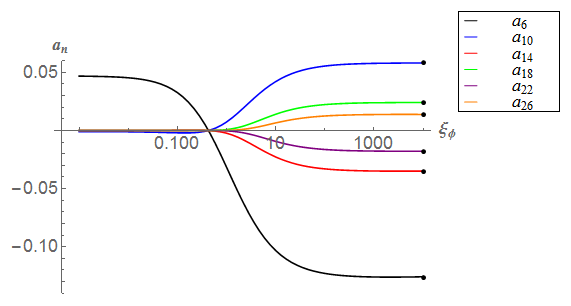}
\caption{ \small \baselineskip 12pt Magnitudes of the first seven nonzero Fourier coefficients $a_n$ of $\phi (\tau)$ as functions of $\xi_\phi$. The $a_n$ have been normalized by the magnitude of the dominant mode $a_2$ for fixed $\xi_\phi$, so $a_2$ is not displayed. The Fourier coefficients corresponding to the asymptotic solution in the $\xi_\phi \rightarrow \infty$ limit are marked by black dots at the right side of the plot. }
\label{anfig}
\end{figure}

For arbitrary $\xi_\phi$, we may expand $\phi (\tau)$ in a Fourier series. The Fourier coefficients $a_n$ for each harmonic are defined by
\beq
a_n = \frac{2}{T} \int_0^T d \tau \> \phi ( \tau ) \cos \left(\frac{n \pi  \tau }{T}  \right) ,
\label{andef}
\eeq
with $T$ given by Eq. (\ref{T}). Given the initial conditions for $\phi (\tau)$ at $\tau_{\rm end}$, the coefficients for all odd $n$ vanish identically. Moreover, as indicated in Eq. (\ref{cncos}) for the $\xi_\phi = 0$ case, all $a_{4n}$ for $n \geq 1$ also vanish --- a feature that remains true (to within numerical precision) for $\xi_\phi \neq 0$. The only nonzero coefficients of the Fourier expansion are $a_{4n+2}$ for $n \geq 0$. In Fig.\ \ref{anfig} we plot the first seven nonzero coefficients $a_n$ as functions of $\xi_\phi$. Consistent with the behavior in Fig.\ \ref{phiFourierfig}, the first two coefficients dominate in the limit $\xi_\phi \sim 0$, whereas a richer spectrum emerges for $\xi_\phi > 1$. Moreover, the dependence of each $a_n$ on $\xi_\phi$ becomes nearly flat for $\xi_\phi \geq 100$, indicating a single asymptotic behavior in the limit $\xi_\phi \gg 1$.
We can analytically prove the existence of an asymptotic solution for the inflaton oscillation, which is independent of $\xi_\phi$ in the limit of $\xi_\phi \to \infty$, if we rescale the field as $\delta \equiv \sqrt{\xi_\phi} \> \phi / M_{\rm pl}$ and time as $\tilde{t} \equiv M_{\rm pl} \, \tau / \xi_\phi = \sqrt{\lambda_\phi} \, M_{\rm pl} \, t / \xi_\phi$. Performing these operations we arrive at the $\xi_\phi$-independent asymptotic equation of motion
\begin{align}
\ddot \delta + {1-\delta^2 \over \delta (1+\delta^2)} \dot \delta^2 + {1 \over 6} {\delta \over 1+\delta^2}=0 ,
\label{eq:asymptoticeq}
\end{align}
Details of the derivation and solution of Eq.\ \eqref{eq:asymptoticeq} are given in Appendix A.

In the limit $\xi_\phi \rightarrow \infty$, for which the background solutions $\delta (t)$ satisfy Eq.~(\ref{eq:asymptoticeq}), we find the Fourier coefficients ($a_2,a_6,a_{10},...$) to be
\beq
 \{a_{4n+2}\} \approx \{{1}, {-0.1265}, \, {0.05813},\, {-0.03513}, \,{0.02415}, \,
{-0.01790}, \,{0.01396}, \, ... \}  \, ,
\eeq
where we have normalized the amplitude of the dominant mode to unity.
We note both an alternation between positive and negative values for rising $n$, as well as a falling magnitude. 
As can be seen in Fig.\ \ref{anfig}, the Fourier coefficients calculated numerically for finite $\xi_\phi$ quickly asymptote to these values once the nonminimal coupling reaches $\xi_\phi \simeq 100$. On the other hand, there remains significant variation of the Fourier coefficients $a_n$ with $\xi_\phi$ for $\xi_\phi \sim {\cal O}(10)$. Whereas the inflationary observables, such as the spectral index $n_s$ and the tensor-to-scalar ratio $r$, attain their large-$\xi_\phi$ values by $\xi_\phi \simeq 10$ \cite{KMS,GKS,KS,SSK}, the post-inflationary dynamics are sensitive to three distinct regimes for $\xi_\phi$, with behavior for $0 < \xi_\phi < 1$ distinct from $\xi_\phi \sim {\cal O} (1 - 10)$, which in turn remains distinct from the regime with $\xi_\phi \geq 100$. For preheating, in other words, there exists an intermediate-$\xi_\phi$ regime, with important consequences for particle production.

For $\xi_\phi > 0$, another significant modification to the oscillation of $\phi (\tau)$ may occur, compared to the $\xi_\phi = 0$ case. In the equation of motion for $\phi$, Eq. (\ref{phieomnoH}), the two distinct contributions from the nontrivial field-space metric --- the noncanonical kinetic term and the conformal stretching of the Einstein-frame potential --- may exactly cancel, resulting in {\it simpler} motion for $\phi (\tau)$ than in the minimally coupled case. In particular, for a specific amplitude $\phi_0 = \tilde{\phi}_0 (\xi_\phi)$, with
\beq
\tilde{\phi}_0 \equiv \frac{ M_{\rm pl} }{\sqrt{ \xi_\phi (6 \xi_\phi - 1 ) } }
\label{tildephi0}
\eeq
and $\xi_\phi > 1/6$, Eq. (\ref{phieomnoH}) reduces to $\phi'' + m^2 \phi = 0$, where primes denote $d / d\tau$ and
\beq
m^2 = \frac{  ( 6 \xi_\phi - 1 ) }{72 \xi_\phi^3 } M_{\rm pl}^2 .
\label{m2magic}
\eeq
 In that case the background field motion is sinusoidal, $\phi (\tau) = \tilde{\phi}_0 \> \cos (m \tau)$, with period 
\beq
T = \frac{2\pi }{m} = \frac{12 \pi}{M_{\rm pl} } \sqrt{ \frac{ 2 \xi_\phi^3 }{6 \xi_\phi - 1} } ,
\label{Tmagic}
\eeq
in exact agreement with the value of $T$ calculated from Eq. (\ref{T}). We thereby find that the inherently anharmonic oscillations in the $\xi_\phi = 0$ case, arising from the nonlinear equation of motion $\phi'' + \phi^3 = 0$, may reduce for $\xi_\phi > 1/6$ to purely harmonic motion for a special value of the amplitude. The fact that there is dependence on the amplitude is not surprising, given that Eq. (\ref{phieomnoH}) is a highly nonlinear equation. 

\begin{figure}
\centering
\includegraphics[width=0.65 \textwidth]{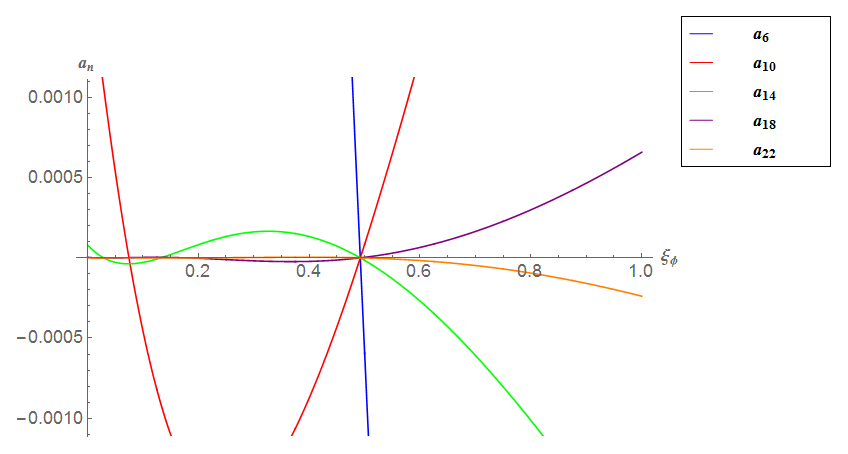}
\caption{ \small \baselineskip 12pt The behavior of the Fourier coefficients $a_{4n+2}$ for $n \geq 1$ within the range $0 \leq \xi_\phi \leq 1$. For a given initial amplitude, $\phi_0$, there exists a single value $\xi_\phi$ at which all higher harmonics vanish identically, leaving purely harmonic evolution for $\phi (\tau)$. }
\label{anfigmagic}
\end{figure}

Given an initial amplitude of oscillation $\phi_0$ at $\tau_{\rm end}$ determined self-consistently from the criterion $\epsilon \equiv - \dot{H} / H^2 = 1$, we may invert the relation between $\tilde{\phi}_0$ and $\xi_\phi$ in Eq. (\ref{tildephi0}) to find the value of $\xi_\phi$ at which all higher harmonics vanish. In Fig.\ \ref{anfigmagic} we plot the Fourier coefficients $a_{4n + 2}$ for $n \geq 1$ in the regime $0 \leq \xi_\phi \leq 1$, and find that all higher harmonics vanish at $\xi_\phi \approx 0.486$. We note, however, that $\tilde{\phi}_0 \sim \xi_\phi^{-1}$ in the limit of large $\xi_\phi$, whereas the amplitude of the inflaton at the end of inflation scales as $\phi_0 \sim \xi_\phi^{-1/2}$, so that the special, simple oscillation of $\phi (t)$ is unlikely to be relevant in the limit $\xi_\phi \gg 1$ before nonlinear effects such as backreaction alter the dynamics.

In summary, we find that the oscillations of $\phi (\tau)$ display richer harmonic structure for $\xi_\phi > 0$ than for the $\xi_\phi = 0$ case. For most values of $\xi_\phi$, we find more non-negligible harmonics than for the minimally coupled case, and the magnitudes of those harmonics rapidly asymptote to fixed values for $\xi_\phi \geq 100$. Meanwhile, for a special amplitude of oscillation, all higher harmonics vanish, again in distinction to the $\xi_\phi = 0$ case. We now turn to the fluctuations, which can be parametrically amplified by the periodic motion of the background field.

\section{Evolution of the Fluctuations}
\label{EvolutionFluctuations}

In Ref. \cite{MultiPreheat1} we established a framework with which to study the evolution of the gauge-invariant fluctuations $Q^I$ during preheating, by expanding the action to second order in both field and metric perturbations, calculating the energy density, and performing a (covariant) mode expansion. This framework will enable us to use Floquet analysis to examine how the resonance structure changes in the presence of $\xi_I \neq 0$ compared to the minimally coupled case.

To first order in the fluctuations, the equation of motion for $Q^I$ may be written \cite{MultiPreheat1}
\beq
{\cal D}_t^2 Q^I + 3 H {\cal D}_t Q^I + \left[ \frac{ k^2}{a^2} \delta^I_{\>\> J} + {\cal M}^I_{\> \>J} \right] Q^J = 0 .
\label{Qeom}
\eeq
Here ${\cal D}_J Q^I = \partial_J Q^I + \Gamma^I_{\> JK} Q^K$ is the covariant derivative with respect to the field-space metric ${\cal G}_{IJ}$, in terms of which the directional derivative may be written ${\cal D}_t Q^I = \dot{\varphi}^J {\cal D}_J Q^I$, where $\varphi^I (t)$ are the spatially homogenous background fields. The mass-squared matrix is given by
\beq
{\cal M}^I_{\>\> J} = {\cal G}^{IK} \left( {\cal D}_J {\cal D}_K V \right) - {\cal R}^I_{\> LMJ} \dot{\varphi}^L \dot{\varphi}^M  - \frac{1}{ M_{\rm pl}^2 a^3} {\cal D}_t \left( \frac{ a^3}{H} \dot{\varphi}^I \dot{\varphi}_J \right) ,
\label{MIJdef}
\eeq
where ${\cal R}_{ILMJ}$ is the Riemann tensor constructed from ${\cal G}_{IJ}$, and the last term in Eq.~(\ref{MIJdef}) arises from the coupled metric perturbations. 

We follow the decomposition of the perturbations using field-space vielbeins, as discussed in detail in Section IV A of Ref. \cite{MultiPreheat1}. We rescale the fluctuations, $Q^I (x^\mu) \rightarrow X^I (x^\mu) / a(t)$, and quantize, $X^I \rightarrow \hat{X}^I$, expanding the fluctuations in creation and annihilation operators and associated mode functions:
\beqn
\begin{split}
\hat{X}^\phi (x^\mu) &= \int \frac{ d^3 k}{(2 \pi)^{3/2} } \left[ \left( v_k e_1^{\> \phi} \hat{b}_{\bf k} + w_k e_2^{\> \phi} \hat{c}_{\bf k} \right) e^{i {\bf k}\cdot {\bf x} } + \left( v_k^* e_1^{\> \phi} \hat{b}_{\bf k}^\dagger + w_k^* e_2^{\> \phi} \hat{c}_{\bf k}^\dagger \right) e^{- i {\bf k}\cdot {\bf x} } \right] , \\
\hat{X}^\chi (x^\mu) &= \int \frac{ d^3 k}{(2 \pi)^{3/2} } \left[ \left( y_k e_1^{\> \chi} \hat{b}_{\bf k} + z_k e_2^{\> \chi} \hat{c}_{\bf k} \right) e^{i {\bf k} \cdot {\bf x} } + \left( y_k^* e_1^{\> \chi} \hat{b}_{\bf k}^\dagger + z_k^* e_2^{\> \chi} \hat{c}_{\bf k}^\dagger \right) e^{- i {\bf k} \cdot {\bf x} } \right] ,
\end{split}
\label{Qphichi}
\eeqn
where the operators obey $\hat{b}_{\bf k} \vert 0 \rangle = \hat{c}_{\bf k} \vert 0 \rangle = 0$ for all ${\bf k}$, and
\beqn
\begin{split}
\left[ \hat{b}_{\bf k} , \hat{b}_{\bf q}^\dagger \right] &= \left[ \hat{c}_{\bf k} , \hat{c}_{\bf q}^\dagger \right] = \delta^{(3)} ( {\bf k} - {\bf q} ) , \\
\left[ \hat{b}_{\bf k} , \hat{c}_{\bf q} \right] &= \left[ \hat{b}_{\bf k} , \hat{c}_{\bf q}^\dagger \right] = 0 .
\end{split}
\eeqn
Within the single-field attractor, ${\cal G}_{\phi \chi} \sim 0$ and hence $e_1^{\> \chi} \sim e_2^{\> \phi} \sim 0$, so that the equations of motion for the mode functions effectively decouple:
\beqn
\begin{split}
v_k'' &+ \Omega_{(\phi)}^2 (k, \eta) \> v_k \simeq 0 , \\
z_k'' &+ \Omega_{(\chi)}^2 (k, \eta) \> z_k \simeq 0 ,
\end{split}
\label{vzeom}
\eeqn
where here primes denote derivatives with respect to conformal time, $d \eta \equiv dt / a(t)$, and the effective frequencies are given by
\beqn
\begin{split}
\Omega_{(\phi)}^2 (k, \eta) &= k^2 + a^2 m_{\rm eff, \phi}^2 (\eta), \\
\Omega_{(\chi )}^2 (k, \eta) &= k^2 + a^2 m_{\rm eff,\chi}^2 (\eta) .
\end{split}
\label{Omegas}
\eeqn
The effective masses may be decomposed into four distinct contributions \cite{MultiPreheat1},
\beq
m_{\rm eff,\phi}^2 = m_{1,\phi}^2 + m_{2, \phi}^2 + m_{3, \phi}^2 + m_{4, \phi}^2 ,
\label{meffphi1}
\eeq
with
\beqn
\begin{split}
m_{1,\phi}^2 &\equiv {\cal G}^{\phi K} \left( {\cal D}_\phi {\cal D}_K V \right) , \\
m_{2, \phi}^2 &\equiv - {\cal R}^\phi_{\>\> LM \phi} \dot{\varphi}^L \dot{\varphi}^M , \\
m_{3, \phi}^2 &\equiv - \frac{1}{ M_{\rm pl}^2 a^3} \delta^\phi_{\>\> I} \delta^J_{\>\> \phi} \> {\cal D}_t \left( \frac{ a^3}{H} \dot{\varphi}^I \dot{\varphi}_J \right) , \\
m_{4, \phi}^2 &\equiv - \frac{ 1}{6} R ,
\end{split}
\label{m1234}
\eeqn
where $R$ is the spacetime Ricci curvature scalar; comparable expressions follow for the contributions to $m_{\rm eff, \chi}^2$. 

Within the rigid-spacetime approximation, $a (t) \rightarrow 1$, $\eta \rightarrow t$, and $m_{3, I}^2, m_{4, I}^2 \rightarrow 0$. Then the only contributions to the fluctuations' effective masses arise from the (covariant) curvature of the potential ($m_{1, I}^2$) and from the curved field-space manifold ($m_{2, I}^2$). (As indicated in Refs.~\cite{MultiPreheat1,Taruya,BKM0,BKM,Sting,FinelliBrandenberger,ShinjiBassett,ChambersRajantie,BondFrolov,Bethke,Moghaddam} and further analyzed in Ref.~\cite{MultiPreheat3}, the contributions from the metric perturbations, $m_{3,I}^2$, may become significant in certain regions of parameter space when one relaxes the rigid-spacetime approximation.)

Since we are considering motion of the background fields within a single-field attractor along the direction $\chi = 0$, $v_k \sim \delta \phi_k$ corresponds to fluctuations in the adiabatic direction, and $z_k \sim \delta \chi_k$ corresponds to fluctuations in the isocurvature direction. The corresponding energy densities are given by \cite{MultiPreheat1}
\beqn
\rho_k^{(\phi)} = {1\over 2} \left ( |\dot v_k|^2 + \Omega^2_{(\phi)}(k,t) |v_k|^2 \right )
\\
\rho_k^{(\chi)} = {1\over 2} \left ( |\dot z_k|^2 + \Omega^2_{(\chi)}(k,t) |z_k|^2 \right ).
\eeqn
We measure particle production with respect to the instantaneous adiabatic vacuum, $\vert 0 (t_{\rm end}) \rangle$, which minimizes the energy densities $\rho_k^{(I)}$ at the end of inflation \cite{AHKK}.

Within the rigid-spacetime approximation, the frequencies $\Omega_{(I)}^2 (k, t)$ oscillate periodically as the background field $\phi (t)$ oscillates periodically. Floquet's theorem then stipulates that solutions to Eq.~(\ref{vzeom}) may be written in the form $v_k (t) = P_1 (k, t) \exp [ \mu_k t ] + P_2 (k, t) \exp [ - \mu_k t ]$ for some periodic functions $P_{1,2}$, and likewise for $z_k (t)$ \cite{AHKK}. The Floquet exponents $\mu_k$ depend, in general, on the couplings, the background oscillation amplitude $\phi_{\rm max}$, and the wavenumber $k$. Regions of parameter space for which ${\rm Re} \> [ \mu_k ] \neq 0$ correspond to exponential instabilities, within which the energy densities $\rho_k^{(I)}$ grow rapidly due to particle production.

In the next two sections, we follow the method of Section 3.2 of Ref. \cite{AHKK} to construct Floquet charts showing regions with ${\rm Re} \> [\mu_k ] \neq 0$ for both adiabatic and isocurvature fluctuations. Whereas calculating the full Floquet chart for this system is a numerical task, semi-analytic progress can be made if one wishes to calculate the boundaries of the instability bands, where the real parts of the Floquet exponents vanish. Since the boundaries of the bands are defined by ${\rm Re} \> [ \mu_k ] =0$, the resulting perturbations $v_k(t)$ and $z_k(t)$ will be periodic there, with period simply related to the periods of the functions $\Omega^2_{(\phi)}(k,t)$ and $\Omega^2_{(\chi)}(k,t)$.

Assuming a rigid spacetime, the only time dependence for the frequencies $\Omega^2_{(I)} (k,t)$ comes from $m_{1, I}^2 (t)$ and $m_{2,I}^2 (t)$, which themselves depend on $\phi (t)$ and $\dot{\phi} (t)$. The periods $T_n$ of $[\phi(t)]^n$ and $[\dot{\phi} (t)]^{n}$ are given by $T$ if $n$ is odd and $T/2$ if $n$ is even, where $T$ is the period of $\phi(t)$. This fact enormously simplifies our consideration of the periods of complicated functions, since in the non-vanishing components of ${\cal M}^I_{\>\> J}$, the only terms that arise are constants and terms of the form $\phi^{2n}(t)$ and $\dot{\phi}^{2n}(t)$, which are all $T/2$ periodic. It follows that each of the relevant components is $T/2$-periodic. Therefore, the period of both $\Omega_{(\phi)}^2$ and $\Omega_{(\chi)}^2$ is half that of the background-field oscillation. We thus look for regions in $(g/{\lambda_\phi}, k^2)$ space that give solutions to Eq.~\eqref{vzeom} that are either $T/2$ or $T$ periodic, since these give the boundaries of stability bands. The details of the method are presented in Appendix B and the results are shown in Section \ref{Isocurvaturemodes}, where they are compared with full numerical calculations of the Floquet charts.

We further note that the exponents $\mu_k$ have units of inverse time, meaning that the magnitude of $\mu_k$ should be defined in comparison to some time-scale. For preheating the relevant time-scale is the Hubble time at the end of inflation. The Hubble time scales like $1/\xi_\phi$ for large $\xi_\phi$ \cite{MultiPreheat1}, hence one relevant parameter that we could plot and compare is $\mu_k \xi_\phi$. This is still a dimensionful quantity, but it has the same parameter dependence as the dimensionless combination $\mu_k / H$. However, one might object to using the Hubble scale as a measure of time in the rigid-spacetime approximation. Instead one may compare the rate $\mu_k$ to the period of background oscillations $T$, and use the quantity $\mu_k  T$. For $\xi_\phi \gg 1$ we found above that $T\sim \xi_\phi$, hence both normalization schemes scale in the same way with $\xi_\phi$: $\mu_k T \sim \mu_k / H \sim \mu_k \xi_\phi$ in the limit $\xi_\phi \gg 1$. (Recall from Section \ref{BackgroundDynamics} that the period $T$ is measured in units of $(\sqrt{\lambda_\phi} \, M_{\rm pl})^{-1}$.) 

Similarly, $k$ is a dimensionful quantity, so we again need to divide by some energy scale in order to find meaningful results. The only energy scale in the problem is the Hubble energy, so the relevant combination is again $k\,  \xi_\phi$. In a static universe, we may consider a characteristic length-scale $l$, defined as $l = c\, T$, which again gives the same relevant scaling for large $\xi_\phi$:  $k  T \sim k\, \xi_\phi$ for $\xi_\phi \gg 1$. Since in the limit of large ${ \xi_\phi}$, the scaling with $T$ and $H$ are equivalent, we will scale the Floquet exponents and wavenumbers by the period of the background oscillation (or simply by $\xi_\phi$) for the remainder of this paper.

It is worth reiterating the applicability and limitation of the Floquet formalism, which requires adopting the rigid-spacetime approximation. For example, one may be concerned about whether such an analysis can accurately describe how long-wavelength modes, with $k \ll aH$, would behave in an expanding universe, since such modes should be sensitive to the expansion of spacetime. However, as shown in Ref.~\cite{KLS}, even the behavior of the $k = 0$ mode and its transition from the broad- to the narrow-resonance regime in an expanding universe may be understood from an analysis of the corresponding (rigid-spacetime) Floquet charts. More generally, modes whose Floquet exponent (as calculated within the rigid-spacetime approximation) satisfies ${\rm Re} [ \mu_k ] \gg H$ will be significantly amplified in an expanding universe. One may estimate the resulting amplification by considering how a given mode would ``flow" through the instability bands of a Floquet chart, as both the mode's physical wavenumber and the amplitude of the oscillating background field redshift. (See, e.g., Refs.~\cite{KLS,AHKK,HertzbergKarouby,MustafaBaumann,Lozanov,MustafaDuration}.) Thus Floquet analysis can often yield both a general intuition as well as a quantitative guide to how the system would behave in a dynamical spacetime, at least for linearized dynamics (before fully nonlinear interactions dominate).

We may now explore the growth rate of perturbations for various values of the nonminimal couplings, potential parameters, and wavenumbers. Since scanning the parameter space by brute force is rather impractical (and wouldn't provide much physical insight), we divide our analysis into subsections, each focusing on a specific range of parameters and type of perturbation (adiabatic or isocurvature). We focus on background-field trajectories within a single-field attractor, and consider (without loss of generality) the attractor to lie along the direction $\chi = 0$. As demonstrated in Ref. \cite{MultiPreheat1}, we may always exploit our covariant framework and perform a rotation in field space so that the attractor lies along the $\chi = 0$ direction.

\section{Results: Adiabatic modes}
\label{Adiabaticmodes}

\begin{figure}
\centering
\includegraphics[width=0.48 \textwidth]{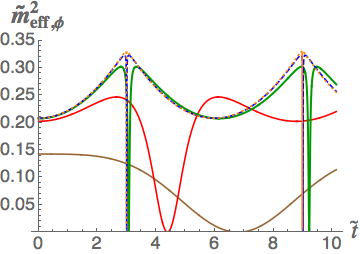}   
\caption{\small \baselineskip 11pt Rescaled effective mass for the adiabatic perturbations, $\tilde{m}_{\rm eff,\phi}^2 = \xi_\phi^2 m_{\rm eff, \phi}^2$ (in units of $M_{\rm pl}$) versus $\tilde{t} \equiv \sqrt{\lambda_\phi} \, M_{\rm pl} \, t / \xi_\phi$, for $\xi_\phi=1,10,10^2,10^3,10^4$ (brown, red, green, blue, and orange, respectively), with $m_{\rm eff,\phi}^2$ given in Eq.~(\ref{eq:meffphi_full}). The sharp features become more pronounced in the limit $\xi_\phi \gg 1$, and asymptote to a single, self-similar behavior. }
\label{adiabaticmasstest1}
\end{figure}

We begin the analysis with the adiabatic modes, which are simpler than the isocurvature modes since their behavior depends only on $\xi_\phi$ and $k$. Within the single-field attractor, the contribution to $m_{\rm eff,\phi}^2$ arising from the curved field-space manifold vanishes, $m_{2,\phi}^2 \sim {\cal O} (\chi\dot{\chi} ) \sim 0$ \cite{MultiPreheat1}. In the rigid-spacetime approximation, therefore, the only contribution to $m_{\rm eff,\phi}^2$ comes from gradients of the potential, $m_{1,\phi}^2$, and may be written
\beq
m_{\rm eff , \phi}^2 = \frac{-2 \delta ^6 (6 \xi_{\phi} +1)+\delta ^4 (12 \xi_{\phi} +1)+3 \delta ^2}{\left(\delta ^2+1\right)^2 \xi_{\phi}  \left(\delta ^2 (6 \xi_{\phi} +1)+1\right)^2} M_{\rm pl}^2,
\label{eq:meffphi_full}
\eeq
where again $\tau = \sqrt{\lambda_\phi} \> t$ and $\delta (\tau) = \sqrt{\xi_\phi} \> \phi (\tau) / M_{\rm pl}$ is the rescaled field amplitude. In the limit $\xi_\phi \gg1$ we may further simplify the effective mass:
\beq
m_{{\rm eff},\phi}^2 \approx  {M_{\rm pl}^2 \over \xi_{\phi}^2}  {(1-\delta^2 ) \over 3 (1+\delta^2)^2} ,
\label{eq:meffphi_approx}
\eeq
where we have set $1 + \xi_\phi \approx \xi_\phi$ as well as $1 + \xi_\phi \delta^n \approx \xi_\phi \delta^n$, since the regime of interest corresponds to $\delta \sim {\cal O} (1)$. (Obviously the second of these approximations fails whenever $\delta (t)$ passes through zero.) If we again rescale time by $\xi_{\phi}$ as $\tilde{t} \equiv M_{\rm pl} \, \tau/\xi_{\phi} = \sqrt{\lambda_\phi} \, M_{\rm pl} \, t / \xi_\phi$, the equation of motion for the adiabatic modes in Eq.~(\ref{vzeom}) takes the form
\beq
\label{eq:adiabaticeomlargexi}
\ddot v_k + ( \tilde k^2 + \tilde m_{{\rm eff},\phi}^2  ) \> v_k=0 ,
\eeq
where $\tilde k = \xi_{\phi} k / (\sqrt{\lambda_\phi} \, M_{\rm pl})$, $\tilde m_{{\rm eff},\phi}^2 = \xi_{\phi}^2 m_{{\rm eff},\phi}^2 / M_{\rm pl}^2$, and overdots denote $d / d \tilde{t}$. As shown in Fig.\ \ref{adiabaticmasstest1} (and also highlighted in Ref.~\cite{Ema}), not only does $\tilde{m}_{\rm eff,\phi}^2 (\tilde{t})$ oscillate as $\phi (\tilde{t})$ oscillates; it also develops sharp features, which become more pronounced for $\xi_\phi \gg 1$, asymptoting to a single, self-similar behavior in the limit $\xi_\phi \rightarrow \infty$. 

\begin{figure}
\centering
\includegraphics[width=0.48 \textwidth]{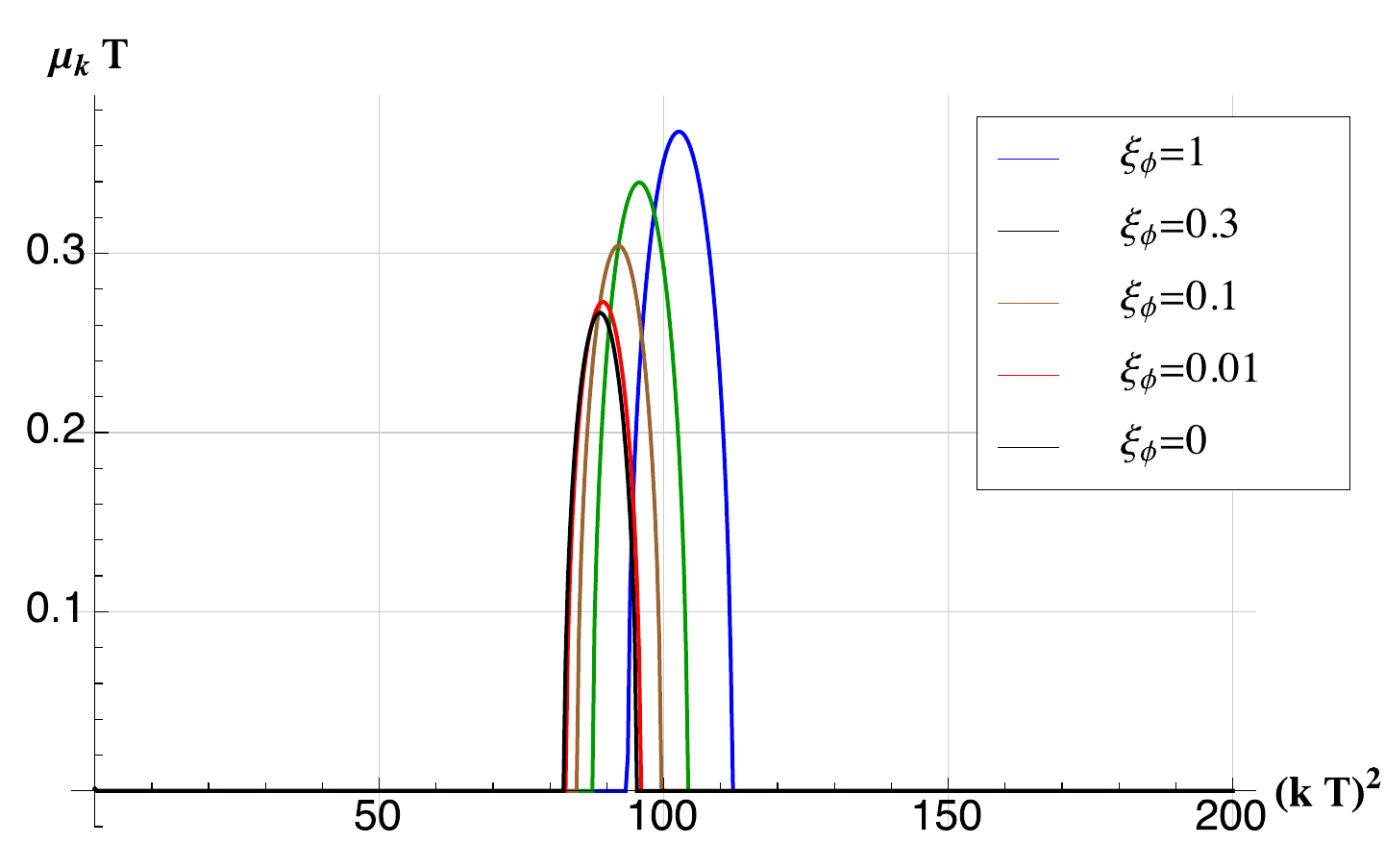} \quad \includegraphics[width=0.48 \textwidth]{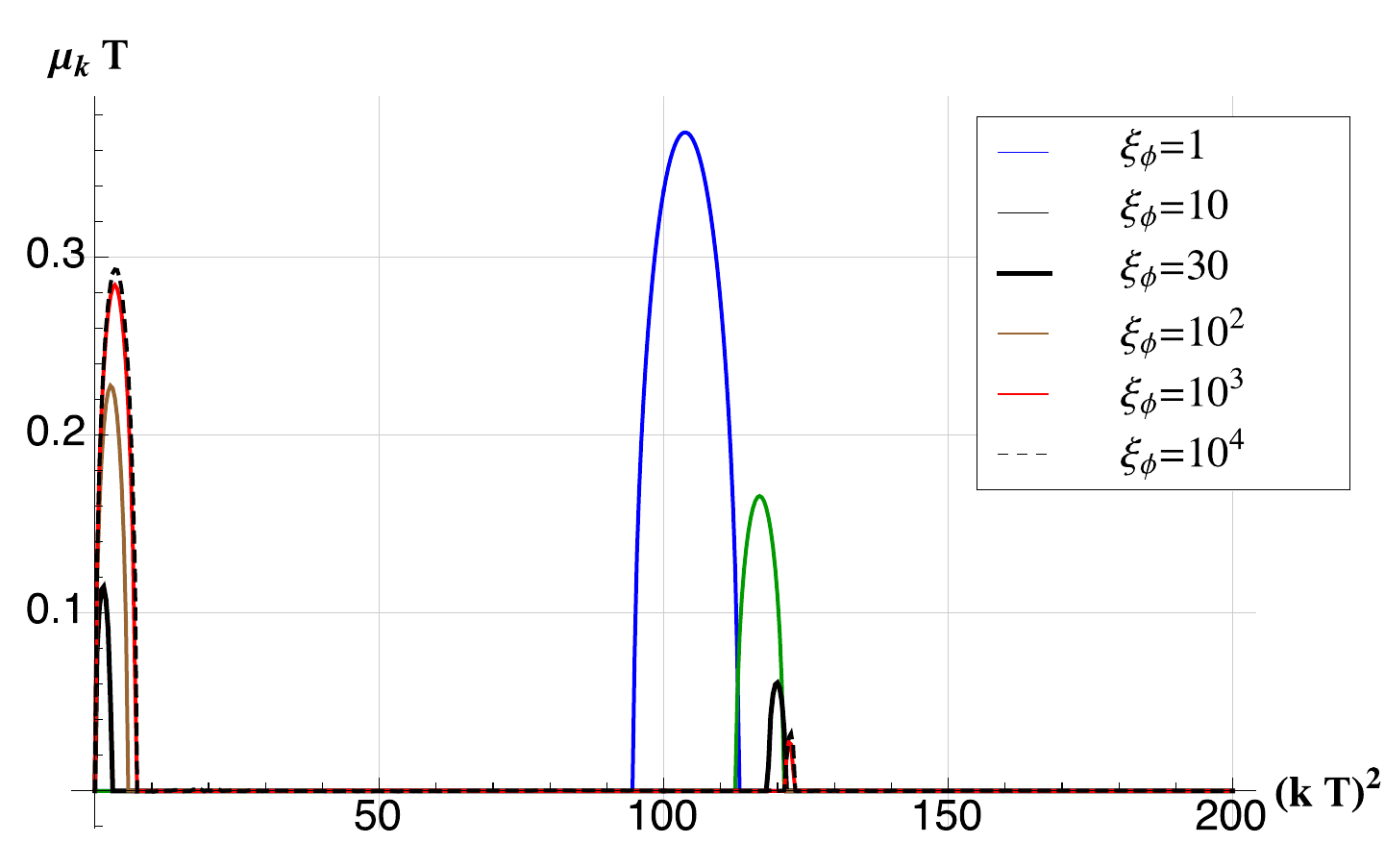}  
\caption{\small \baselineskip 11pt ({\it Left}) The normalized Floquet exponent ${\rm Re} [\mu_k] T$ for the adiabatic mode $v_k$ for $0 \leq \xi_\phi \leq 1$. ({\it Right}) The normalized Floquet exponent ${\rm Re} [\mu_k] T$ for the adiabatic mode $v_k$ for $\xi_\phi \geq 1$. The common behavior for $\xi_\phi \gg 1$ is evident.  }
\label{fig:muk_vk}
\end{figure}

As expected, the time dependence of $\tilde{m}_{\rm eff, \phi}^2 (\tilde{t})$ drives parametric resonances in the adiabatic modes $v_k (\tilde{t})$. Fig.\ \ref{fig:muk_vk} summarizes the dependence of the (real part of the) Floquet exponent on the wavenumber and the nonminimal coupling. Starting with the large-$\xi_\phi$ regime, where interesting self-similar behavior can be found, we see that the Floquet chart for $\xi_\phi \gg 1$ asymptotes to a common shape, once $\mu_k$ and $k$ are rescaled by $\xi_\phi$, as explained in Section \ref{EvolutionFluctuations}. The large-$\xi_\phi$ scaling behavior may be understood analytically: after rescaling time and wavenumber by $\xi_{\phi}$ in Eq. \eqref{eq:adiabaticeomlargexi}, $\xi_\phi$ drops out of the equation altogether.

In the small-$\xi_\phi$ regime, the dominant Floquet band occurs at a nonzero wavenumber. As $\xi_\phi$ increases from 0 to 1, the band moves to larger wavenumbers and the magnitude of ${\rm Re} [\mu_k ] T$ is enhanced by as much as $30\%$. For $\xi_\phi \sim 1$, a new band emerges at $k\sim 0$. As $\xi_\phi$ increases further and we move into the large-$\xi_\phi$ regime, the band at $k\sim 0$ becomes the dominant band. For asymptotically large $\xi_\phi$, the value of ${\rm Re} [ \mu_k ] T$ for the dominant band at $k\sim 0$ becomes about $6$ times larger than for the secondary instability band.

\begin{figure}
\centering
\includegraphics[width=0.6 \textwidth]{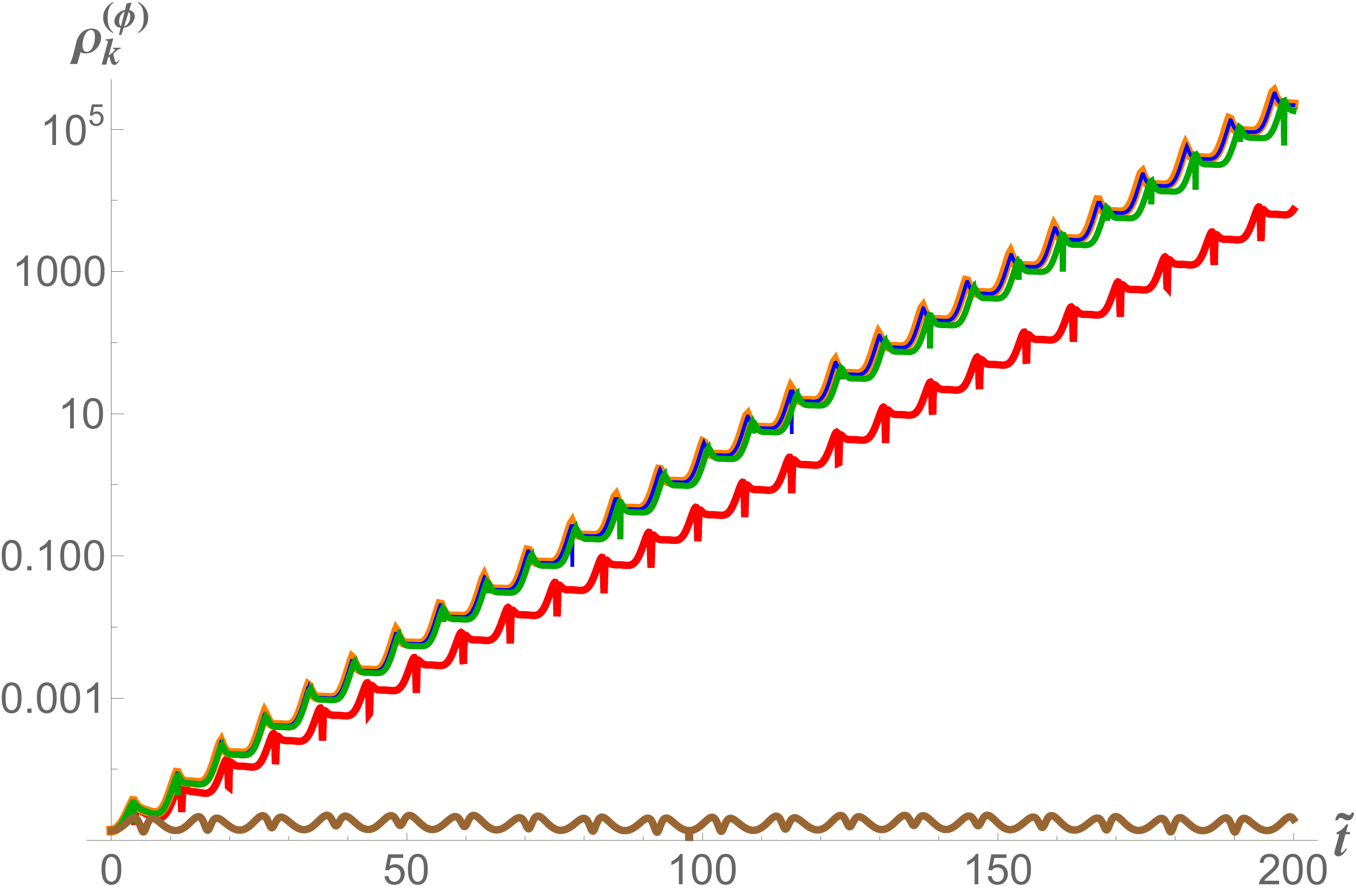}
\caption{\small \baselineskip 11pt Energy density $\rho^{(\phi )}_k$ versus $\tilde{t} = \sqrt{\lambda_\phi} \, M_{\rm pl} \, t / \xi_\phi$ for $\tilde{k}^2 = [\xi_\phi k / (\sqrt{\lambda_\phi} \, M_{\rm pl} )]^2 = 0.03$ and $\xi_\phi = 1, 10, 10^2, 10^3, 10^4$ (brown, red, green, blue, and orange, respectively). The energy density rapidly asymptotes to a single behavior in the limit of large $\xi_\phi$. }
\label{fig:adiabaticfixedk}
\end{figure}

In Fig.~\ref{fig:adiabaticfixedk} we plot $\rho^{(\phi)}_k$ for various values of $\xi_\phi$ and for the wavenumber $\tilde{k}^2 = 0.03$. Consistent with the scaling of the Floquet exponents shown in Fig.~\ref{fig:muk_vk}, we find three distinct regimes for $\xi_\phi$ at these long wavelengths: no growth for $\xi_\phi = 1$, modest growth for $\xi_\phi = 10$, and then a quick approach to a single behavior of rapid growth for $\xi_\phi \geq 10^2$. (Because we are neglecting nonlinear effects like the backreaction of created particles on the evolution of the background field, the exponential growth of $\rho^{(\phi)}_k$ within a given resonance band appears to continue forever. Of course when nonlinear effects are incorporated, the resonant amplification will end after a characteristic time \cite{BKM,AHKK,MustafaDuration,Figueroa}, though such effects are beyond the scope of the present study.)

\begin{figure}
\centering
 \includegraphics[width=.49\textwidth]{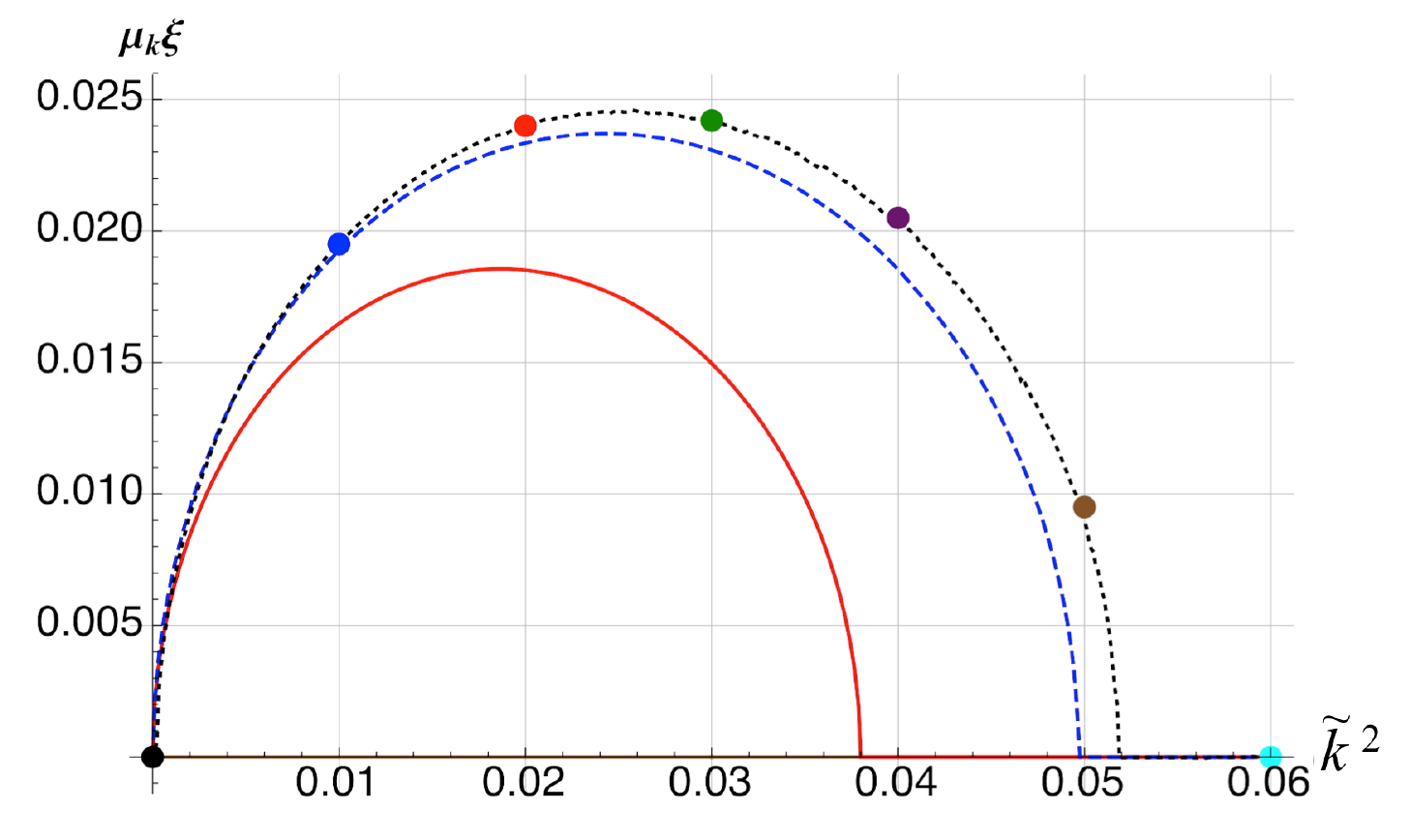} \quad \includegraphics[width=.48\textwidth]{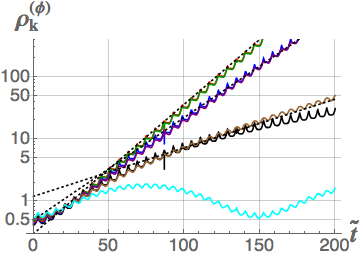}
\caption{\small \baselineskip 11pt
({\it Left}) The Floquet chart for adiabatic perturbations with $\xi_\phi=10^2,10^3,10^4$ (red, blue, and black, respectively). The colored dots correspond to the rescaled wavenumbers $\tilde{k} = \xi_\phi k / (\sqrt{\lambda_\phi} \, M_{\rm pl} )$ used in the right panel. 
({\it Right}) Energy density for the adiabatic mode with $\xi_{\phi}=10^4$ and $\tilde{k}^2 = 0,\, 0.01, \,0.02, \,0.03, \,0.04, \,0.05, 0.06$ (black, blue, red, green, purple, brown, and cyan, respectively). The black-dotted lines correspond to exact exponential growth, $\exp ( 2 {\rm Re} [\mu_k] t )$. Note that the Floquet chart ({\it left}) is normalized with respect to $\xi_\phi$, to make the comparison with the evolution of the energy density as a function of $\tilde{t} = \sqrt{\lambda_\phi} \, M_{\rm pl} \,  t/\xi_\phi$ ({\it right}) easier.}
\label{fig:multi_k}
\end{figure}

In Fig.\ \ref{fig:multi_k}, we zoom in on modes $\tilde{k} \ll 1$, which lie within the dominant Floquet instability band in the limit $\xi_\phi \gg 1$. On the righthand side we plot the energy density of the adiabatic modes. The black-dotted lines correspond to exact exponential growth at the corresponding value of the Floquet exponent, $\exp ( 2 {\rm Re} [\mu_k] t)$; they match very well with the late-time exponential growth of $\rho^{(\phi)}_k$ at various wavenumbers. The behavior of the $k=0$ mode seems quite paradoxical at first. According to the Floquet chart, ${\rm Re} [\mu_{k = 0} ] = 0$. We would therefore expect the mode to be non-growing, or at most oscillating like the $\tilde{k}^2 =0.06$ mode. 
However we see that the growth is not exponential, but rather polynomial, with $v_k(\tilde{t}) \propto \tilde{t}$, which is the growth exhibited by systems having two zero eigenvalues. Hence the growth of the $k=0$ mode is consistent with the Floquet chart.

\begin{figure}
\centering
 \includegraphics[width=.48\textwidth]{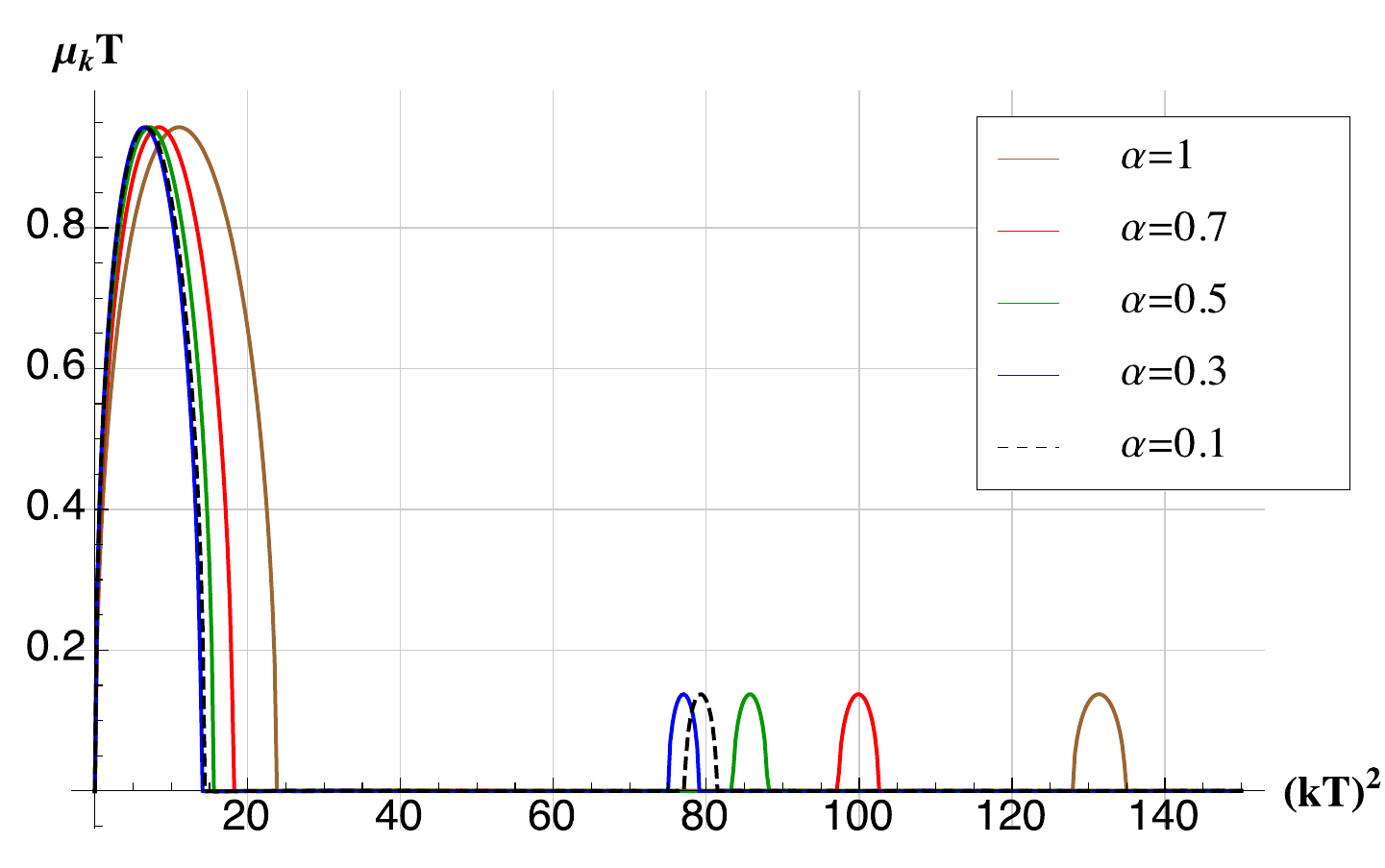} \quad \includegraphics[width=.48\textwidth]{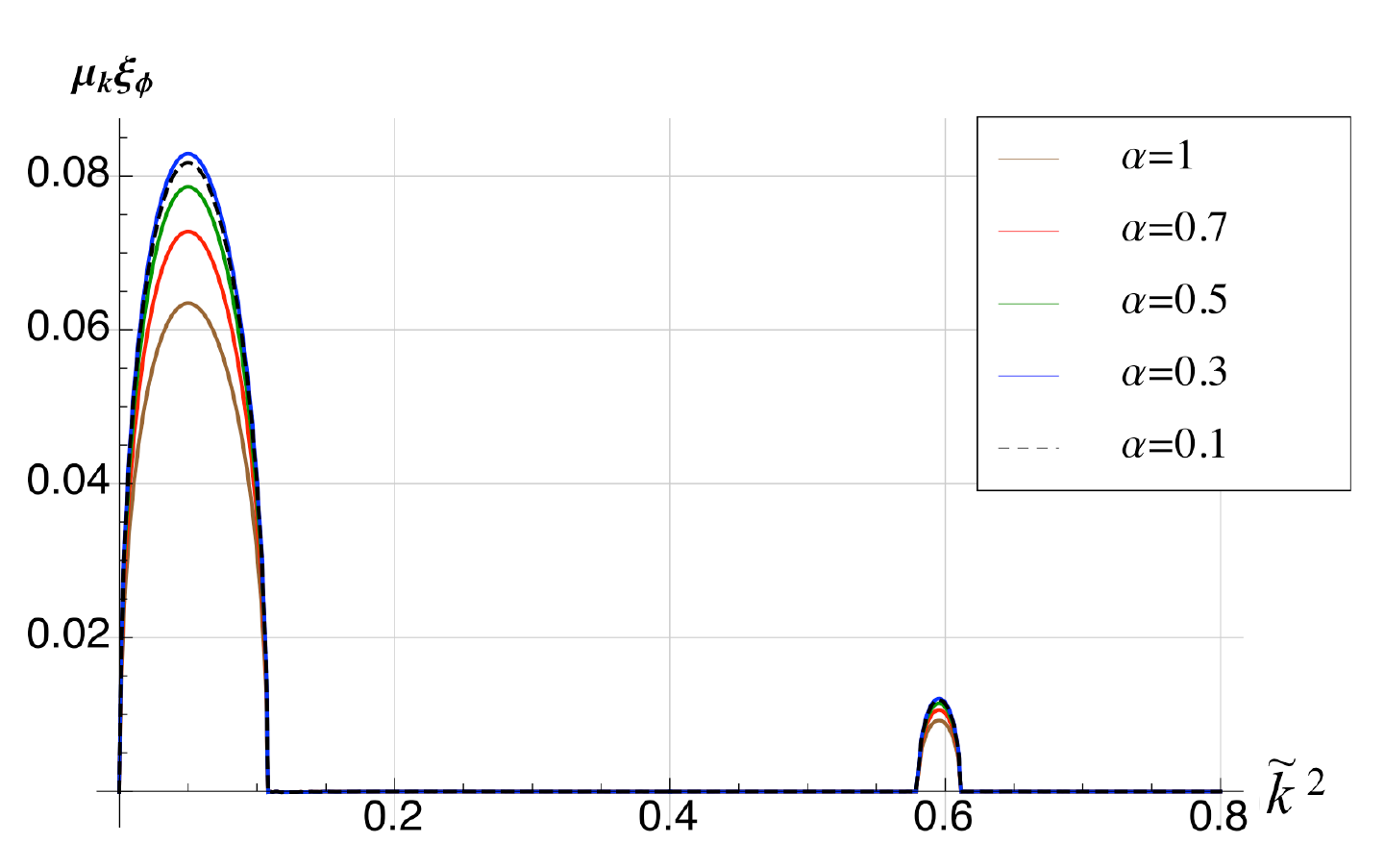}
\caption{\small \baselineskip 11pt
The Floquet chart for adiabatic perturations with $\xi_\phi = 10^3$, and the amplitude of the background oscillation parameterized as $\delta(0) = (\alpha\times 0.8) \sqrt{\xi_\phi} \> \phi_0 / M_{\rm pl}$, with varying $\alpha$. ({\it Left}) The Floquet exponent versus wavenumber, each scaled by the period of background oscillations $T$. ({\it Right}) The Floquet exponent scaled by the nonminimal coupling $\xi_\phi$, versus the rescaled wavenumber $\tilde{k} = \xi_\phi k / (\sqrt{ \lambda_\phi} \, M_{\rm pl} )$.  }
\label{fig:floquet_alpha}
\end{figure}

Next we consider the effect of the background amplitude parameter $\alpha$ on the amplification of adiabatic modes. The Floquet chart for large $\xi_\phi$ and varying values of $\alpha$ is shown in Fig.\ \ref{fig:floquet_alpha}. An interesting phenomenon occurs depending on the choice of normalization. If one normalizes the Floquet exponent and wavenumber by $T$, then each band has a constant amplitude for different values of $\alpha$, but differs in width and position. However, if one normalizes by $\xi_\phi$, then each band has a fixed position and width, but its amplitude varies as a function of $\alpha$. Using $\xi_\phi$ as a normalization parameter is physically justified if we use the appropriate factors of $M_{\rm pl}$ to construct a dimensionless quantity.

In sum, we identified a scaling solution for the growth of adiabatic modes at large $\xi_\phi$. The Floquet chart accurately predicts the late-time growth rate of the adiabatic modes. As one increases $\xi_\phi$, the Floquet chart approaches an asymptotic form. The most important difference between the large-$\xi_\phi$ resonance structure and the corresponding structure for the minimally coupled case is the emergence of a dominant instability band around $k \sim 0$ for large nonminimal coupling. The fact that the dominant band corresponds to $k \sim 0$ means that overall amplification should be increased, compared to the minimally coupled case: physical modes in an expanding universe will redshift {\it toward} the resonance band rather than out of it. In the meantime, special care is required for handling the $k = 0$ mode, which grows linearly in a static universe.

\section{Results: Isocurvature modes}
\label{Isocurvaturemodes}

We now proceed to the case of isocurvature perturbations, where the dependence of $\mu_k$ on $g/\lambda_\phi$ and $\xi_\phi / \xi_\chi$ provides room for richer phenomenology. Since we want to satisfy the attractor condition along the $\chi=0$ direction, we will choose $g/\lambda_{\phi} \geq 1$, which, as shown in Section III A of Ref. \cite{MultiPreheat1}, is a sufficient condition for a potential valley along that direction for $\xi_\phi = \xi_\chi$ and any nonzero value of $\xi_\phi$. 

\begin{figure}
\centering
\includegraphics[width=0.48 \textwidth]{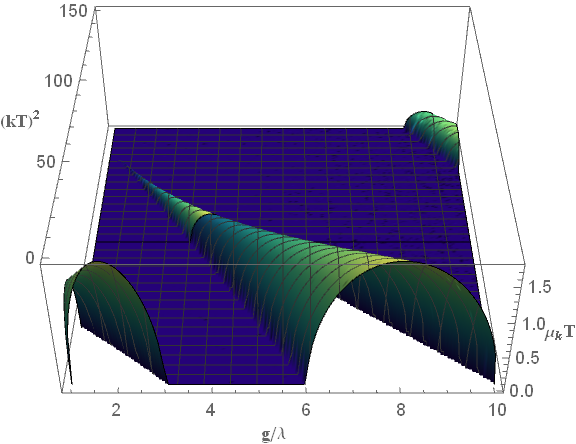} \quad     \includegraphics[width=0.48 \textwidth]{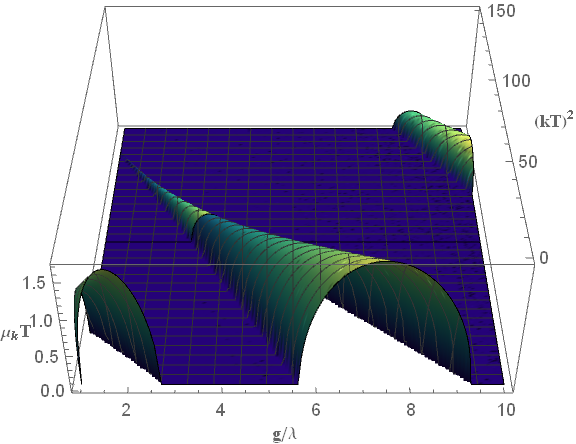}  
\\
\includegraphics[width=0.48 \textwidth]{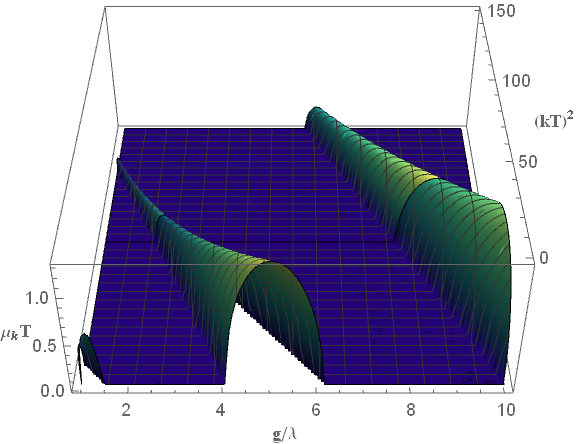} \quad  \includegraphics[width=0.48 \textwidth]{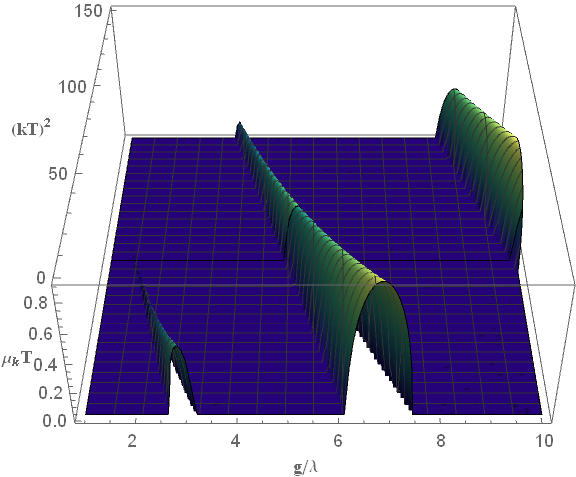}  
\\
\includegraphics[width=0.48\textwidth]{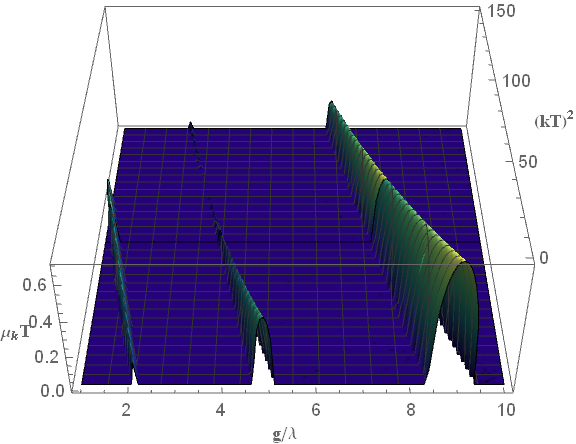} \quad  \includegraphics[width=0.48 \textwidth]{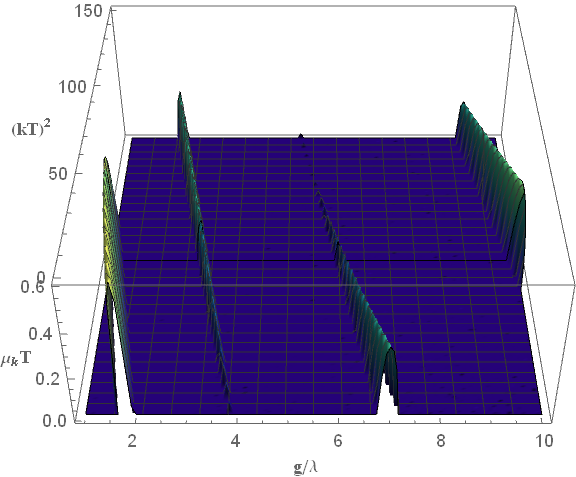}  
\caption{\small \baselineskip 11pt Normalized Floquet exponents ${\rm Re} [ \mu_k ] T$ for the isocurvature perturbations $z_k$ as functions of $g / \lambda_\phi$ and $(kT)^2$, for $\xi_\phi = \xi_\chi$. ({\it Top, left to right}) $\xi_\phi = 0, 0.01$; ({\it middle, left to right}) $\xi_\phi = 0.1, 0.4$; ({\it bottom, left to right}) $\xi_\phi = 0.7, 1$. The bands are visibly tilting, shifting, and squeezing with increasing nonminimal coupling, indicating less efficient preheating for $\xi_\phi \sim {\cal O} (1)$ compared to the $\xi_\phi = 0$ case. }
\label{fig:zxi3D_low_xi}
\end{figure}

We start with $\xi_\phi \le 1$. As shown in Fig.\ \ref{fig:zxi3D_low_xi}, the results for $\xi_\phi=0$ reproduce the familiar Lam\'e chart for the minimally coupled case \cite{Boyanovsky,KaiserPreh2,GKLS}. As we increase $\xi_\phi$, the bands get shifted to lower values of the coupling $g/\lambda_{\phi}$, and tilted further away from the $g/\lambda_{\phi}$ axis. The regions with ${\rm Re} [\mu_k] \neq 0$ become reduced in width and height. In other words, in the regime $0 < \xi_\phi \leq 1$, preheating is less efficient than in the minimally coupled case.

\begin{figure}
\centering
\includegraphics[width=0.48 \textwidth]{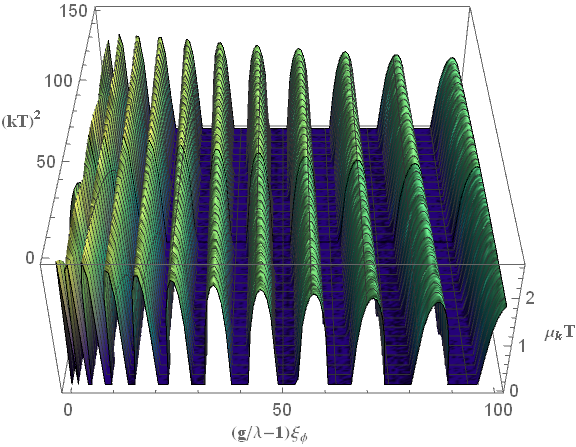} \quad   \includegraphics[width=0.48 \textwidth]{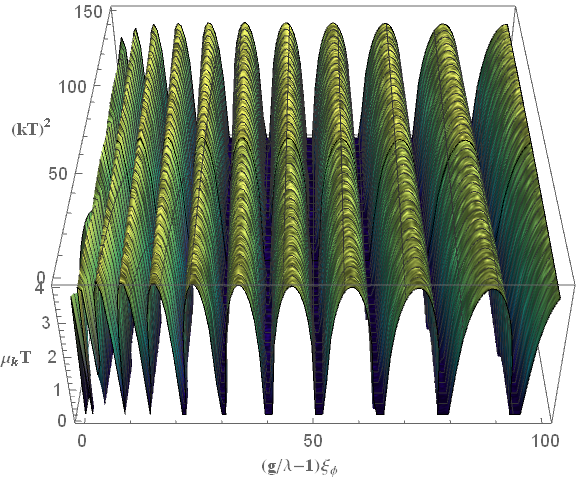}  
\\
\includegraphics[width=0.48 \textwidth]{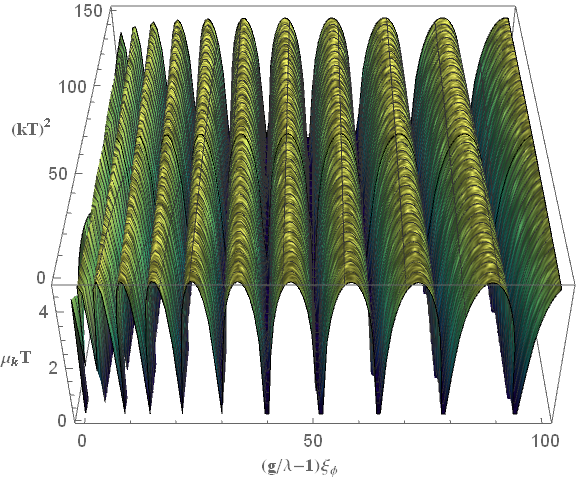}  \quad \includegraphics[width=0.48 \textwidth]{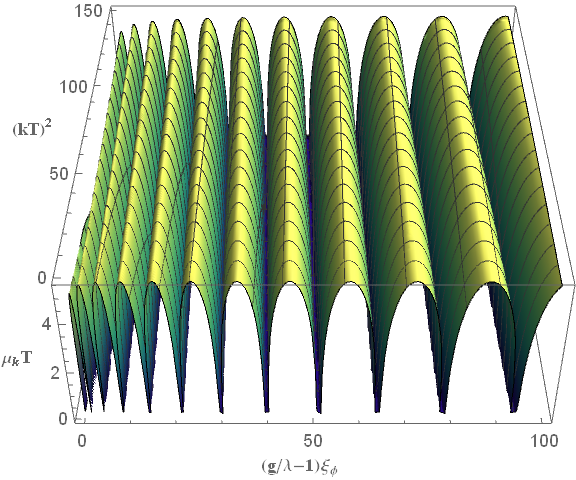}  
\caption{\small \baselineskip 11pt Normalized Floquet exponents ${\rm Re} [\mu_k] T$ for the isocurvature perturbations $z_k$, as functions of $[ (g / \lambda_\phi ) - 1 ] \xi_\phi$ and $(kT)^2$, for $\xi_\phi = \xi_\chi$. ({\it Top, left to right}) $\xi_\phi = 10, 10^2$; ({\it bottom, left to right}) $\xi_\phi = 10^3, 10^4$. The similarity of the Floquet charts for large $\xi_\phi$ is visible.}
\label{fig:zxi3D_large_xi}
\end{figure}

This situation is reversed in the large-$\xi_\phi$ regime. In order to study this regime we perform one further rescaling, by the dimensionless quantity $g/\lambda_{\phi}$. As discussed in Ref.~\cite{MultiPreheat1}, for large $\xi_\phi$ the direction $\chi = 0$ will be an attractor in field space whenever $\tilde{\Lambda}_\phi < 0$, where
\beq
\tilde{\Lambda}_\phi \equiv \frac{\xi_\chi}{\xi_\phi} - \frac{g}{\lambda_\phi} .
\label{tildeLambdadef}
\eeq
The strength of the attractor is governed by the combination $\tilde \Lambda_\phi \xi_\phi$; hence the attractor gets stronger for larger values of $\xi_\phi$. For a non-elliptical potential, in which the nonminimal couplings are equal $(\xi_\phi = \xi_\chi$), the attractor-strength parameter may be re-written $\tilde \Lambda_\phi \xi_\phi = - [ (g/\lambda_{\phi}) -1]\xi_\phi$. Since the strength of the attractor is a characterization of the curvature of the potential in the adiabatic direction, we will use the combination $[(g / \lambda_\phi) - 1] \xi_\phi$ as the effective coupling strength, rather than $g/\lambda_\phi$, when examining the Floquet charts for large $\xi_\phi$. Fig.\ \ref{fig:zxi3D_large_xi} shows the 3-dimensional Floquet charts, in which one may see a convergence into a forest of densely packed, large-valued, almost parallel instability bands for large $\xi_\phi$. We find substantially more efficient amplification of isocurvature modes for $\xi_\phi \gg 1$ than for the $\xi_\phi = 0$ case.

\begin{figure}
\centering
\includegraphics[width=0.48\textwidth]{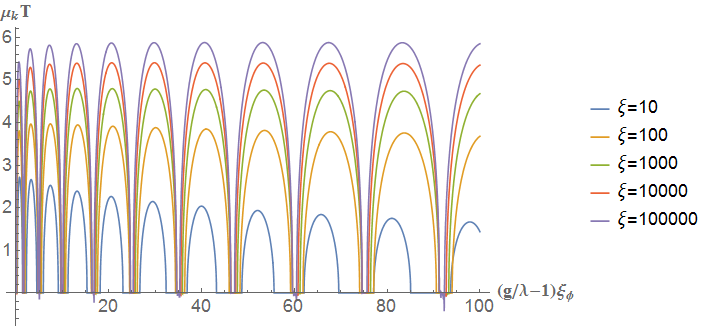}  \quad \includegraphics[width=0.48 \textwidth]{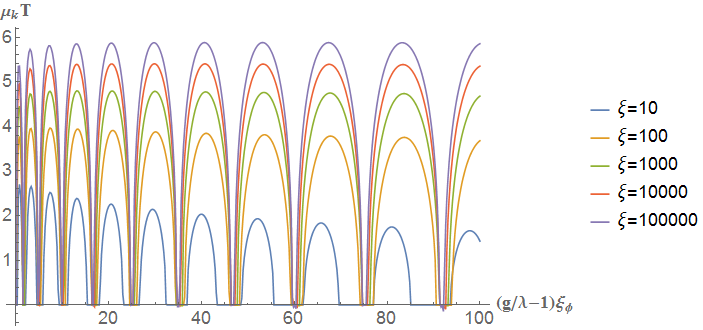}  
\\
\includegraphics[width=0.48 \textwidth]{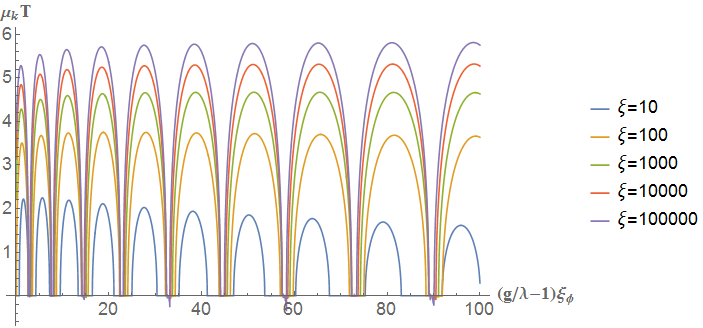} \quad  \includegraphics[width=0.48\textwidth]{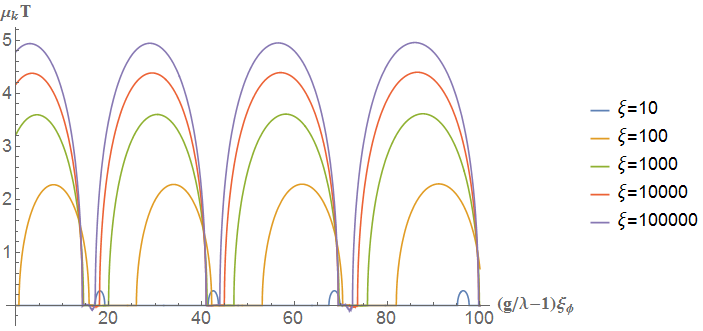}  
\caption{\small \baselineskip 11pt Normalized Floquet exponents ${\rm Re} [\mu_k] T$ for the isocurvature perturbations $z_k$, as functions of $[ (g / \lambda_\phi) - 1] \xi_\phi$ and $(kT)^2$, for $\xi_\phi = \xi_\chi = 10,10^2,10^3,10^4,10^5$. Each two-dimensional plot corresponds to a slice through the full Floquet chart of Fig.\ \ref{fig:zxi3D_large_xi} at a specific normalized wavenumber: ({\it top, left to right}) $kT= 0,1$, ({\it bottom, left to right}) $kT = 10,100$. }
\label{fig:zxi1D_large_xi}
\end{figure}

In order to more readily examine the structure of these Floquet charts, we present 2-dimensional slices of them in Fig.\ \ref{fig:zxi1D_large_xi}, in which each panel shows the Floquet exponent for a specific rescaled wavenumber $kT$ as a function of the effective coupling $[(g/\lambda_\phi) -1] \xi_\phi$. There are several interesting points about the band structure in the regime of large nonminimal couplings. First, there is an increase of the value of ${\rm Re} [ \mu_k] $ as one increases the nonminimal coupling. The approach to an asymptotic solution is not as immediate as in the adiabatic case, though the $\xi_\phi=10^4$ case differs from the $\xi_\phi=10^5$ case only by about $10\%$. For $\xi_\phi=10$ the value of ${\rm Re} [\mu_k]$ is higher at lower effective coupling. This situation is reversed as one increases the nonminimal coupling: for large $\xi_\phi$, nearly all bands have essentially the same height. Finally, from Fig.\ \ref{fig:zxi1D_large_xi} we note that the bands tilt and become narrower for increasing $k$, though the effect only becomes pronounced for $kT \geq 10$. 

In the large-$\xi_I$ regime, we thus find a distinct behavior of the structure of the instability bands that has no analogue in models with minimally coupled fields. Whereas our goal for this analysis has been to characterize the resonance structure, and note significant differences from previous, well-studied models, the behavior shown in Figs.~\ref{fig:zxi3D_large_xi} and \ref{fig:zxi1D_large_xi} raises additional, interesting questions. In particular, as discussed in Ref.~\cite{BezrukovRunning}, the perturbative unitarity scale $\Lambda$ for these models becomes a function of the inflaton amplitude. In the Einstein frame, in the limit $\xi_\phi \gg 1$, the strong-coupling scale $\Lambda$ scales as $M_{\rm pl}$ during inflation (for $\phi \geq M_{\rm pl} / \sqrt{\xi_\phi}$) before asymptoting to $\Lambda \sim M_{\rm pl} / \xi_\phi$ for small field values ($\phi < M_{\rm pl} / \xi_\phi$). Therefore one should take care that the wavenumbers $k$ under consideration do not extend into the regime $k > \Lambda$. 

Within the rigid-spacetime approximation (which we adopt throughout this paper), the strong-coupling regime does not enter the analysis, since $\Lambda \propto M_{\rm pl} \rightarrow \infty$; hence {\it every} mode with finite comoving wavenumber satisfies $k \ll \Lambda$. Moreover, even if we restore $M_{\rm pl}$ to its usual value, we find at the start of preheating, when $\phi \sim M_{\rm pl} / \sqrt{\xi_\phi}$ and $\Lambda \sim M_{\rm pl}$, that $k < \Lambda$ for $kT < 14.8 \, \xi_\phi$, given that the inflaton period $T$ scales as $T \simeq 14.8 \, \xi_\phi / M_{\rm pl}$ in the large-$\xi_I$ regime \cite{MultiPreheat1}. Later during preheating, as the amplitude of $\phi$ falls and $\Lambda \rightarrow M_{\rm pl } / \xi_\phi$, we expect modes with $kT \lesssim 10$ to satisfy $k < \Lambda$. Of course, to fully explore the dynamics of the system beyond the linearized analysis we have pursued here, one would also need to consider particle rescatterings, from which large $k$ modes could enter the strong-coupling regime. An analysis of the high-$k$ behavior of these models beyond linear order in the perturbations remains beyond the scope of the present paper, and we leave potential implications of the strong-coupling regime for future research.

In sum, for the isocurvature modes $z_k$ and symmetric nonminimal couplings ($\xi_\phi = \xi_\chi$), there is a weak convergence with increasing $\xi_\phi$ to an asymptotic solution that has large values of ${\rm Re} [\mu_k]$ across dense, almost-parallel instability bands. The dense instability profile for the isocurvature modes for large $\xi_\phi$ is related to the rich spectral content of the background field for large $\xi_\phi$, as identified in Figs.~\ref{phiFourierfig} and \ref{anfig}. We may make sense of the behavior shown in Figs.~\ref{fig:zxi3D_low_xi}-\ref{fig:zxi1D_large_xi} semi-analytically.

\subsection{Small $\xi_\phi$}

As shown in Fig.~\ref{fig:zxi3D_low_xi}, preheating into isocurvature modes becomes {\it less} efficient for $0 < \xi_\phi < 1$ compared to the minimally coupled case, with $\xi_\phi = 0$. We may understand this trend qualitatively by returning to Fig.~\ref{anfig}, which shows the amplitudes of the first few nonzero Fourier coefficients $a_n$ of the background field's evolution. In the limit $\xi_\phi \rightarrow 0$, only the first two harmonics ($a_2$ and $a_6$) have any sizable amplitude, and the ratio between them is quite large, with $a_2 / a_6 \sim 20$. As one approaches $\xi_\phi \rightarrow 1$ from below, one still finds only these two dominant harmonics, but the magnitude of $a_6$ falls relative to $a_2$, such that the background field's oscillations become more nearly sinusoidal. With fewer nontrivial harmonics in the background field's evolution, the coupled perturbations experience fewer resonances, and the overall amplification falls. For $\xi_\phi > 1$, on the other hand, several harmonics begin to rise in magnitude, yielding the richer and more efficient resonance structure depicted in Figs.~\ref{fig:zxi3D_large_xi} and \ref{fig:zxi1D_large_xi}.  

We may further understand properties of the Floquet charts by examining the Fourier structure of certain field-space quantities. In the rigid-spacetime limit, Eq.~(\ref{vzeom}) for the isocurvature modes $z_k$ may be written in the suggestive form
\beq
{d \over dt} \begin{pmatrix}
z_k \\
\dot{z}_k
\end{pmatrix} =  \begin{pmatrix}
0 & 1 \\
-(k^2 + m_{{\rm eff},\chi}^2) & 0
\end{pmatrix} \begin{pmatrix}
z_k \\
\dot{z}_k
\end{pmatrix} , 
\label{eqn:vielbeinmatrix}
\eeq
again using $m_{\rm eff,\chi}^2 = m_{1,\chi}^2 + m_{2,\chi}^2$ in the rigid-spacetime limit. This equation is of the form 
\beq
\dot{x}(t) = A(t) \> x(t) \, ,
\label{eq:1stordereq}
\eeq
where $A(t)$ is a periodic matrix with period $T/2$, and $T$ is the period of the background-field oscillation. This allows us to employ the machinery of Floquet theory to semi-analytically study the boundaries that separate resonant from non-resonant regions in parameter space. In particular, at the boundaries between stable and unstable regions (at which ${\rm Re} [\mu_k] = 0$), there exist $T/2$-periodic and $T$-periodic solutions for $z_k (t)$.

We describe the method in Appendix B and summarize the results here. The stability boundaries for the isocurvature modes $z_k$ are given implicitly by the equations 
\beqn
\begin{split}
\det \left[ Z((kT)^2, g/\lambda_\phi) \right] &= 0
\\
 \det \left[ \mathcal{Z}((kT)^2, g/\lambda_\phi) \right] &=  0
\\
\det \left[ Z'((kT)^2, g/\lambda_\phi) \right] &= 0
\\
\det \left[ \mathcal{Z}'((kT)^2, g/\lambda_\phi) \right] &= 0 \, ,
\end{split}
\label{ZZprimeeqs}
\eeqn
where the matrices $Z$, $\mathcal{Z}$, $Z'$ and $\mathcal{Z}'$ are functions of $g/\lambda_\phi$ and $k$. For example, the components of the matrix $Z$ are given by
\beqn
\begin{split}
Z_{00} &= k^2 + b_{k,0} . \\
Z_{p,p} &= - 4 p^2 \omega^2 + k^2 + b_{k,0} + \frac{1}{2} b_{k, 2p} \>, \>\>\> p \geq 1 , \\
Z_{p, q} &= \frac{1}{2} \left( b_{k, \vert p - q \vert } + b_{k, p + q} \right) , \>\>\> p \neq q ,
\label{Zcomponents}
\end{split}
\eeqn
where $\omega = 2 \pi / T$ and $b_{k, p}$ are the coefficients of a (cosine) Fourier expansion of $\Omega_{(\chi)}^2 (k, t)$. The related $Z'$, ${\cal Z}$, and ${\cal Z}'$ matrices are constructed in Appendix B. There are four matrices altogether for the isocurvature modes, arising from sine and cosine Fourier expansions for both the $T/2$- and $T$-periodic solutions.

Solving the equations in Eq.~(\ref{ZZprimeeqs}) would give exact (implicit) equations for the boundaries of all stable regions, but --- since these matrices are infinite dimensional --- doing so is computationally intractable. However, by truncating these matrices at an appropriate order we may understand the origin of the band-tilting with increasing $\xi_\phi$. In particular, as demonstrated in Appendix B, truncating to $3 \times 3$ matrices provides sufficiently accurate approximations with which to understand the shift of the boundaries of the fundamental instability band. (For higher-order instability bands, one may truncate the matrices to sizes larger than $3 \times 3$, though understanding the behavior of the fundamental instability band will suffice for our purposes.)

For small $\xi_\phi$, the boundaries of the regions ${\rm Re} [\mu_k] = 0$ are approximately linear in the $(kT)^2$-versus-$g / \lambda_\phi$ plane, which means that the slope is approximately constant everywhere along a given boundary. The intercepts along the $(kT)^2$-axis are the solutions to the determinant equations of Eq.~(\ref{ZZprimeeqs}) with $g/\lambda_\phi = 0$, and those along the $g/\lambda_\phi$-axis are the solutions with $(kT)^2 = 0$. Since all of the relevant matrices depend only on the Fourier coefficients of $m_{{\rm eff},\chi}^2$, the change in slope of the ${\rm Re} [\mu_k] = 0$ boundary in the $(kT)^2 - (g / \lambda_\phi)$ plane --- which we call ``band tilting" --- may be determined analytically by calculating the Fourier coefficients of $d m_{{\rm eff},\chi}^2 / d \xi_\phi$. 

\begin{figure}
\centering
\includegraphics[width=0.48\textwidth]{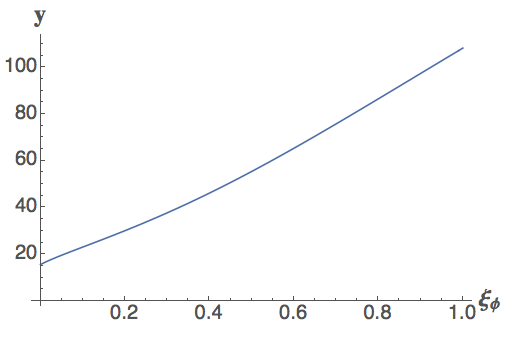} \quad \includegraphics[width=0.48\textwidth]{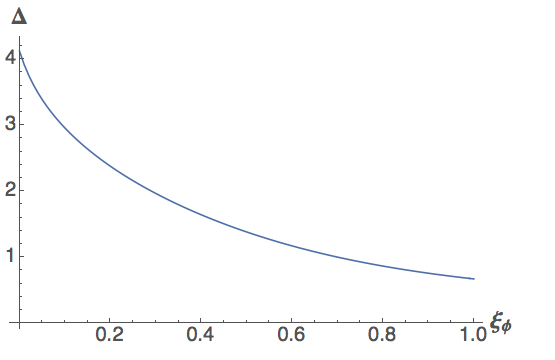}  
\caption{\small \baselineskip 11pt ({\it Left}) Slope $y$ of the upper boundary of the primary stability band as a function of $\xi_\phi$ for $0 \leq \xi \leq 1$. This plot shows that stability bands tilt upward in the $(kT)^2 - (g / \lambda_\phi)$ plane as $\xi_\phi$ increases. ({\it Right}) The width of the primary instability band $\Delta$ as a function of $\xi_\phi$, where $\Delta$ is defined as the difference between the $g/\lambda_\phi$-intercepts of the upper and lower stability boundaries. }
\label{fig:band_tilt}
\end{figure}

We write the contributions to $m_{\rm eff, \chi}^2$ in a suggestive form, in which we isolate the dependence on the coupling parameter $g$: 
\beqn
\begin{split}
m_{{1},\chi}^2 &= \Bigg\{ \frac{g \delta^2}{\xi_\phi (\delta^2 + 1)^2}  -\frac{\delta^4 \left(\delta ^2 (6 \xi_\phi +1)+6 \xi_\phi +2\right)}{\left(\delta ^2+1\right)^3 \xi_\phi  \left(\delta ^2 (6 \xi_\phi +1)+1\right)} \Bigg\} M_{\rm pl}^2,
\\
m_{{2},\chi}^2 &= \frac{2 \dot \delta^2  \left(\delta ^2 (6 \xi_\phi +1)+3 \xi_\phi +1\right)}{\left(\delta ^2+1\right)^2 \left(\delta ^2 (6 \xi_\phi +1)+1\right)} M_{\rm pl}^2,
\label{eq:m2chi}
\end{split}
\eeqn
where again we use $\delta \equiv \sqrt{ \xi_\phi} \> \phi / M_{\rm pl}$. In order to isolate the band-tilting effect, we study one particular stability band, which we call the \emph{primary stability band}: the large central band in the first Floquet chart in Fig.~\ref{fig:zxi3D_low_xi}. We consider how the upper stability boundary shifts with changing $\xi_\phi$. (Similar results may be obtained for the other stability boundaries.) For $\xi_\phi = 0$, the upper stability boundary has a slope in the $(kT)^2$ - $g / \lambda_\phi$ plane of about 15; as $\xi_\phi$ increases toward 1, the slope rises sharply to roughly 100, as shown on the left in Fig.~\ref{fig:band_tilt}. This band-tilting effect is consistent with the features displayed in the Floquet charts of Fig.~\ref{fig:zxi3D_low_xi}.

Another effect displayed in the Floquet charts in Fig.~\ref{fig:zxi3D_low_xi} is that the bands become narrower as $\xi_\phi$ increases. Again we focus on the primary instability band, and consider its upper and lower boundaries. The width of the band, $\Delta$, can be quantified by the difference between the $g/\lambda_\phi$-intercepts of the upper and lower stability boundaries. The narrowing of $\Delta$ with increasing $\xi_\phi$ is shown on the right in Fig.~\ref{fig:band_tilt}, which again agrees with the results of Fig.~\ref{fig:zxi3D_low_xi}.

\subsection{Large $\xi_\phi$}

For large $\xi_\phi$, we again consider Eqs.~(\ref{eqn:vielbeinmatrix}) and (\ref{eq:m2chi}), but now examine how the distinct contributions to $m_{\rm eff, \chi}^2$ scale with $\xi_\phi$. We begin with $m_{{2},\chi}^2$ and consider the behavior for $\xi_\phi \gg 1$ and $\delta \ne 0$:
\beq
\frac{m_{{2},\chi}^2}{M_{\rm pl}^2}  = \left (  \frac{2 \delta ^2+1}{\delta ^2 \left(\delta ^2+1\right)^2}  + {\cal O} (\xi_{\phi}^{-1}) \right )  \dot \delta^2 \sim {1\over \xi_{\phi}^2} ,
\label{m2chilarge}
\eeq
since $\dot \delta$ scales like $1/\xi_{\phi}$ if we change the time variable to $t\to \tilde{t} = \sqrt{\lambda_\phi} \, M_{\rm pl} \, t / \xi_{\phi}$. Using the same dimensional arguments for $m_{{1},\chi}^2$ we find
\beq
\frac{m_{{1},\chi}^2}{M_{\rm pl}^2} = \frac{\delta ^2 [(g / \lambda_\phi ) -1] }{\left(\delta ^2+1\right)^2 \xi_{\phi} } + {\cal O} (\xi_{\phi}^{-2}) \, .
\label{m1chilarge}
\eeq
Eqs.~(\ref{m2chilarge}) and (\ref{m1chilarge}) reveal different scaling for the two components of $m_{\rm eff, \chi}^2$, which makes the analysis of the isocurvature modes in the large-$\xi_\phi$ regime more difficult than for the adiabatic modes. Furthermore, the two components of $m_{\rm eff, \chi}^2$ oscillate out of phase, since one is proportional to $\delta^2$ and the other to $\dot{\delta}^2$. Their distinct scaling with $\xi_\phi$ controls the features of $m_{\rm eff, \chi}^2$, such as when it has sharp features and when it crosses zero. (See also Refs.~\cite{MultiPreheat1,Ema}.)

We demonstrated in Ref.~\cite{MultiPreheat1} that the attractor behavior along the $\chi = 0$ direction is controlled by the combination $\tilde{\Lambda}_\phi \xi_\phi$, which (as we saw in Figs.~\ref{fig:zxi3D_large_xi} and \ref{fig:zxi1D_large_xi}) is also the effective coupling that makes the Floquet charts exhibit self-similar scaling behavior in the large-$\xi_\phi$ limit. For the case $\xi_\phi = \xi_\chi$ (zero ellipticity), we may perform a change of variables
\beq
\frac{g}{\lambda_\phi} = 1 - \tilde{\Lambda}_\phi = 1 + \frac{ a_\phi}{\xi_\phi} ,
\label{aphidef}
\eeq
where we have introduced the parameter $a_\phi \equiv - \tilde{\Lambda}_\phi \xi_\phi$. (Recall that the attractor along $\chi = 0$ corresponds to $\tilde{\Lambda}_\phi < 0$; here $a_\phi$ is chosen to denote the word ``attractor," and should not be confused with either the scale factor $a (t)$ or the Fourier coefficients $a_n$.) The parameter $a_\phi$ is proportional to the parameter $\kappa \equiv 4 (\lambda_\phi \xi_\chi - g \xi_\phi ) / \lambda_\phi$ defined in Ref.~\cite{SSK}, which was introduced to study the evolution of isocurvature perturbations during inflation along the top of a ridge in the potential; here we will use $a_\phi$ to study preheating of isocurvature modes within a valley of the potential.

Using the attractor parameter $a_\phi$, the term $m_{1,\chi}^2$ in Eq.~(\ref{eq:m2chi}) may be written
\beq
\frac{m_{{1},\chi}^2 }{M_{\rm pl}^2 }  = \frac{\delta ^2 \left(\xi_{\phi} -a_\phi  \left(\delta ^2+1\right) \left(\delta ^2 (6 \xi_{\phi} +1)+1\right)\right)}{\left(\delta ^2+1\right)^3 \xi_{\phi} ^2 \left(\delta ^2 (6 \xi_{\phi} +1)+1\right)} = \frac{1-6 a_\phi  \left(\delta ^4+\delta ^2\right)}{6 \left(\delta ^2+1\right)^3 \xi_{\phi} ^2}  +{\cal O}( \xi_{\phi}^{-3}).
\eeq
We can therefore distinguish three distinct regimes of parameters relevant to the analysis of $m_{\rm eff,\chi}^2$ for the isocurvature perturbations, all within the limit of zero ellipticity ($\xi_\phi = \xi_\chi$): a symmetric potential, with $g = \lambda_\phi$ and hence $a_\phi = 0$; a softly broken symmetry, with $\vert \tilde{\Lambda}_\phi \vert \sim \xi_\phi^{-1} \ll 1$ and hence $a_\phi \sim {\cal O} (1)$; and a generic potential, with $\vert \tilde{\Lambda}_\phi \vert \sim {\cal O} (1)$ and hence $a_\phi \sim \xi_\phi$. For $a_\phi \sim {\cal O} (1)$, we find $m_{1,\chi}^2 \sim m_{2, \chi}^2 \sim \xi_\phi^{-2}$, whereas for $a_\phi \gg 1$ we have $m_{1,\chi}^2 \sim \xi_\phi^{-1}$ and $m_{2,\chi}^2 \sim \xi_\phi^{-2}$.

\subsubsection{Symmetric Potential, $a_\phi = 0$}
\label{sec:symmetricpotential}

\begin{figure}
\centering
\includegraphics[width=0.48\textwidth]{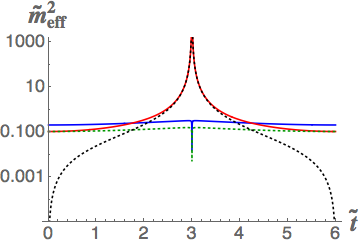}  
 \caption{\small \baselineskip 11pt
The rescaled effective masses, $\tilde{m}_{\rm eff}^2 = \xi_\phi^2 m_{\rm eff}^2$, for the adiabatic (blue) and isocurvature (red) modes (in units of $M_{\rm pl}$) versus $\tilde{t} = \sqrt{\lambda_\phi} \, M_{\rm pl} \, t / \xi_\phi$, for a symmetric potential ($\xi_\phi = \xi_\chi$, $g = \lambda_\phi = \lambda_\chi$), with $\xi_\phi = 10^3$. The green and black dashed curves show the contributions to $\tilde{m}_{\rm eff,\chi}^2$ arising from the potential ($\tilde{m}_{1,\chi}^2$) and from the nontrivial field-space manifold ($\tilde{m}_{2,\chi}^2$), respectively. 
}
\label{fig:AdiabIsoc}
\end{figure}

An interesting case is that of a symmetric potential, with $\xi_\phi = \xi_\chi$ and $g = \lambda_\phi = \lambda_\chi$ (and hence $a_\phi = 0$). Apart from its simplicity, the symmetric case is relevant to models like Higgs inflation \cite{BezrukovShaposhnikov,GKS}. (For the Higgs inflation case there are $3$ identical isocurvature modes \cite{GKS}, but this does not change our analysis.) Fig.~\ref{fig:AdiabIsoc} shows the behavior of the rescaled effective masses $\tilde{m}_{\rm eff}^2 = \xi_\phi^2 m_{\rm eff}^2$ for the adiabatic and isocurvature modes, for $\xi_\phi \gg 1$. We note that the sharp features in $\tilde{m}_{\rm eff,\phi}^2$ that we identified in Fig.~\ref{adiabaticmasstest1} are swamped by the spikes in $\tilde{m}_{\rm eff,\chi}^2$ in the limit $\xi_\phi \gg 1$ (a trend also noted in Ref.~\cite{Ema}). In particular, the spikes of $\tilde{m}_{\rm eff,\chi}^2$ for $\xi_\phi \gg 1$ arise from the nontrivial field-space manifold ($\tilde{m}_{2,\chi}^2$) rather than from the potential ($\tilde{m}_{1,\chi}^2$), and hence have no analogue in the minimally coupled case.

\begin{figure}
\centering
\includegraphics[width=.48\textwidth]{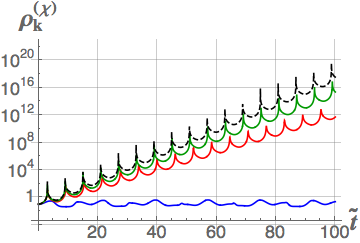}  \quad \includegraphics[width=.48\textwidth]{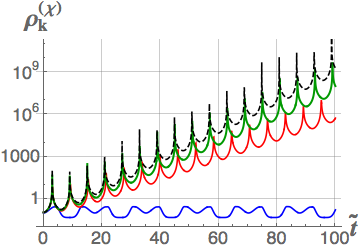}  
 \caption{
\small \baselineskip 11pt Energy density $\rho^{(\chi)}_k$ for the isocurvature modes $z_k$ in a symmetric potential ($\xi_\phi = \xi_\chi$, $g = \lambda_\phi = \lambda_\chi$), as a function of $\tilde{t} = \sqrt{\lambda_\phi} \, M_{\rm pl} \, t / \xi_\phi$. The wavenumbers are $\tilde{k} = 0$ ({\it left}) and $\tilde{k} = 0.1$ ({\it right}), with $\tilde{k} = k\xi_\phi / (\sqrt{\lambda_\phi} \, M_{\rm pl} ).$ In both plots, the nonminimal couplings are given by $\xi_\phi =  10,10^2,10^3,10^4$ (blue, red, green, black-dashed respectively). Compare with the growth of adiabatic modes shown in Figs.~\ref{fig:adiabaticfixedk} and \ref{fig:multi_k}. }
\label{fig:nvnzSYM}
\end{figure}

The energy density of the isocurvature modes likewise quickly exceeds that of the adiabatic modes in the large-$\xi_\phi$ limit. In Fig.~\ref{fig:nvnzSYM} we plot $\rho^{(\chi)}_k$ for the case of symmetric couplings, for various values of $\xi_\phi$ and two distinct modes $k$. We note first the sensitivity to wavenumber: both modes shown here become amplified exponentially quickly, but the $k = 0$ mode grows even more quickly than the $k \xi_\phi = 0.1$ mode. For both modes, we find no particular growth for $\xi_\phi = 10$, and then a convergence toward a single large-$\xi_\phi$ behavior for $\xi_\phi \geq 100$. Compare with the more modest amplification of the adiabatic modes for $\xi_\phi \gg 1$ shown in Figs.~\ref{fig:adiabaticfixedk} and \ref{fig:multi_k}. (As in our analysis of the adiabatic modes, we neglect nonlinear effects such as backreaction; our goal is to understand the earliest phases of the preheating resonances.)

\subsubsection{Softy Broken Symmetry, $a_\phi \sim {\cal O}(1)$}
\label{smoothvalley}

\begin{figure}
\centering
  \includegraphics[width=.48\textwidth]{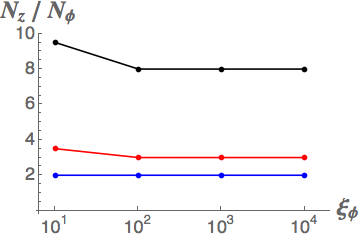}
 \caption{\small \baselineskip 11pt
The number of times the isocurvature mode $z_k (t)$ with $k = 0$ oscillates per background oscillation ($N_z / N_\phi$) as a function of $a_\phi$ and $\xi_\phi$, where $g / \lambda_\phi = 1 + (a_\phi / \xi_\phi)$: $a_\phi = 2$ (blue), $a_\phi = 20$ (red), and $a_\phi = 200$ (black).}
\label{fig:isoevol}
\end{figure}

For the case of a softy broken symmetry, with $\vert \tilde{\Lambda}_\phi \vert \ll 1$ and hence $a_\phi \sim {\cal O} (1)$, we find that the isocurvature modes $z_k (t)$ oscillate more times per background oscillation for larger values of $a_\phi$, as shown in Fig.~\ref{fig:isoevol}. This means that the system can move from the narrow to the broad resonance regime \cite{KLS}, in which the (normalized) effective mass of the isocurvature modes becomes large. The system exits the regime of softly broken symmetry once $a_\phi$ becomes comparable to $\xi_\phi$.

\begin{figure}
\centering
\includegraphics[width=.48\textwidth]{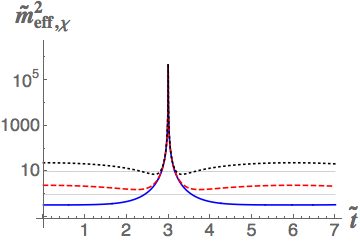}\quad \includegraphics[width=.48\textwidth]{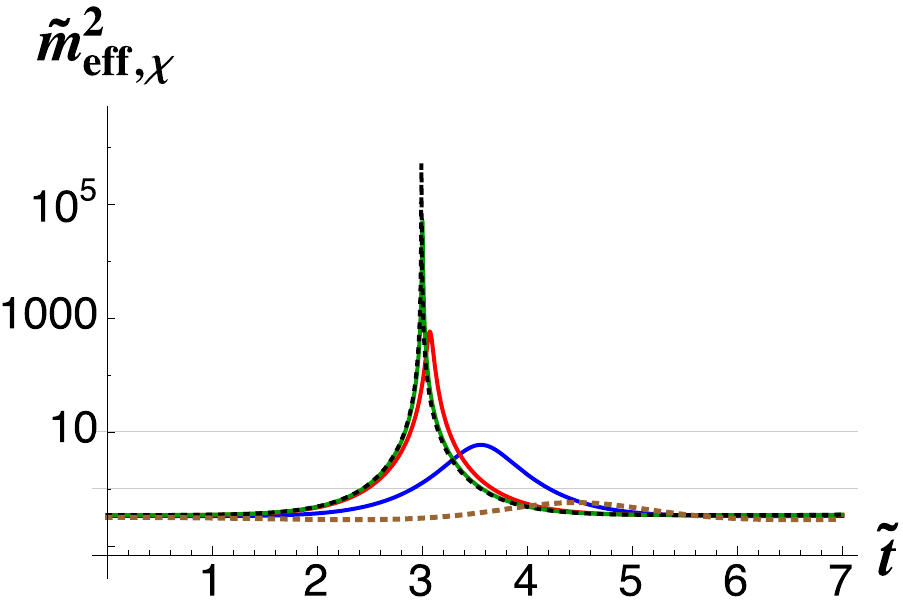} 
 \caption{\small \baselineskip 11pt
({\it Left}) The rescaled effective mass for the isocurvature modes $\tilde{m}_{\rm eff,\chi}^2 = \xi_\phi^2 m_{\rm eff,\chi}^2$ (in units of $M_{\rm pl}$) versus $\tilde{t} = \sqrt{\lambda_\phi} \, M_{\rm pl} \, t / \xi_\phi$ for a softly broken symmetric potential with $\xi_{\phi} = \xi_\chi =10^4$ and $g / \lambda_\phi =1+ (2/\xi_\phi)$ (blue), $g/ \lambda_\phi =1+ (20/\xi_\phi)$ (red), and $g / \lambda_\phi =1+ (200/\xi_\phi)$ (black). ({\it Right}) The effective mass $\tilde{m}_{\rm eff,\chi}^2$ for $g/\lambda_\phi =1+ (2/\xi_\phi)$ and $\xi_{\phi}=3, 10,10^2,10^3,10^4$ (brown, blue, red, green, and black, respectively).  }
\label{fig:meffvara}
\end{figure}

\begin{figure}
\centering
\includegraphics[width=.48\textwidth]{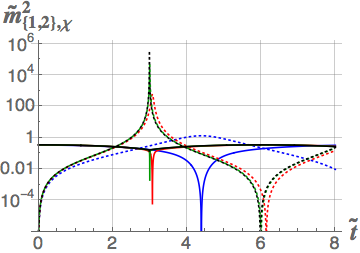} \quad \includegraphics[width=.48\textwidth]{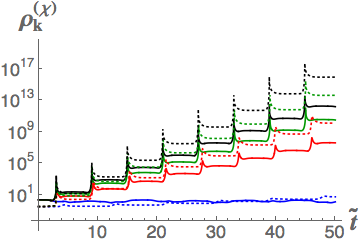} 
\caption{\small \baselineskip 11pt
({\it Left}) Contributions $\tilde{m}_{ \{ 1,2 \}, \chi}^2 = \xi_\phi^2 m_{ \{ 1,2\} , \chi}^2$ to $\tilde{m}_{\rm eff,\chi}^2$ versus $\tilde{t} = \sqrt{\lambda_\phi} \, M_{\rm pl} \, t / \xi_\phi$ for a softly broken symmetric potential with $\xi_\phi = \xi_\chi$,  $g / \lambda_\phi = 1 + (a_\phi / \xi_\phi)$, and $a_\phi = 2$. Solid lines correspond to the contributions that arise from the potential ($\tilde{m}_{1,\chi}^2$) and dotted lines correspond to the contributions that arise from the nontrivial field-space manifold ($\tilde{m}_{2,\chi}^2$). ({\it Right}) The energy density $\rho^{(\chi)}_k$ for $k = 0$ and $\xi_\phi = \xi_\chi$, with $g / \lambda_\phi = 1 + (a_\phi / \xi_\phi)$ and $a_\phi = 2$ (dotted) and $a_\phi = 20$ (solid). In both plots, the nonminimal couplings are given by $\xi_{\phi}=10,10^2,10^3,10^4$ (blue, red, green, and black, respectively).  }
\label{fig:misolargexi}
\end{figure}

Fig.\ \ref{fig:meffvara} shows the isocurvature effective mass for $\xi_{\phi}=10^4$ and varying $a_\phi$ (or equivalently varying $g / \lambda_\phi$). As expected, the spike does not change, since it arises from the nontrivial field-space manifold $(\tilde{m}_{2,\chi}^2$) rather than from the potential ($\tilde{m}_{1,\chi}^2$), and hence is independent of $g / \lambda_\phi$. The potential contribution $\tilde{m}_{1,\chi}^2$ grows with growing $g/ \lambda_\phi$, leading the system from narrow to broad resonance as $\tilde{m}_{{\rm eff},\chi}$ becomes larger than the background frequency. By keeping $a_\phi$ fixed and varying $\xi_\phi$, we see that the spike in the isocurvature effective mass becomes more pronounced for larger values of the nonminimal coupling. In the case of $\xi_\phi=10$ there is no sharp feature, since the two components of the isocurvature effective mass are similar in magnitude and opposite in phase. We thus see again the clear distinction between the intermediate- and large-$\xi_\phi$ regime in the context of preheating. These trends are reinforced in Fig.~\ref{fig:misolargexi}, which depicts the different contributions to $\tilde{m}_{\rm eff,\chi}^2$ from $\tilde{m}_{1,\chi}^2$ and $\tilde{m}_{2,\chi}^2$, and the dependence of $\rho_k^{(\chi)}$ on $a_\phi$ and $\xi_\phi$ for the softly broken symmetry case.

\subsubsection{Arbitrary valley, $a_\phi \sim \xi_\phi$}
\label{arbitraryvalley}

We now consider a generic potential, meaning that $g / \lambda_\phi \sim {\cal O} (1)$ has some arbitrary value (different from $g / \lambda_\phi = 1$, which corresponds to the symmetric case). We will use $g/ \lambda_\phi=2$ for definiteness. In this case, $a_\phi \sim \xi_\phi \gg 1$.

\begin{figure}
\centering
\includegraphics[width=0.48\textwidth]{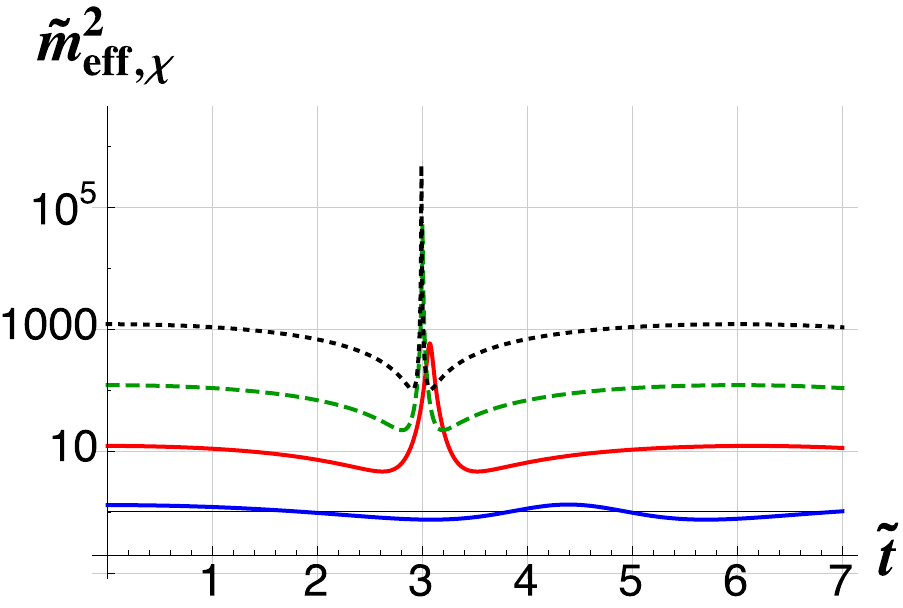}
 \caption{ \small \baselineskip 11pt
The rescaled effective mass $\tilde{m}_{\rm eff,\chi}^2$ versus $\tilde{t} = \sqrt{\lambda_\phi} \, M_{\rm pl} \, t / \xi_\phi$ for $g / \lambda_\phi = 2$ (or $a_\phi = \xi_\phi$), and $\xi_{\phi}=10,10^2,10^3,10^4$ (blue, red, green, and black, respectively). }
\label{fig:meffrandomridge}
\end{figure}

The isocurvature effective mass is shown in Fig.\ \ref{fig:meffrandomridge} for a range of nonminimal couplings. We can see that the overall magnitude of $\tilde{m}_{\rm eff,\chi}^2$ increases with increasing $\xi_\phi$ for a fixed value of $g/\lambda_\phi$. This is easily understood. For $\delta \ne 0$ the velocity is $\dot \delta \sim 0$, hence the potential term dominates: $\tilde{m}_{1,\chi}^2 \gg \tilde{m}_{2,\chi}^2$. For $a_\phi \sim \xi_\phi$, we found above that $m_{1,\chi}^2 \sim \xi_\phi^{-1}$, and hence $\tilde{m}_{1,\chi}^2 = \xi_\phi^2 m_{1,\chi}^2 \sim \xi_\phi$, consistent with the behavior shown in Fig.~\ref{fig:meffrandomridge} away from the spike. Thus larger $\xi_\phi$ values lead to larger masses for the isocurvature perturbations. On the other hand, for $\xi_\phi \sim {\cal O} (10)$, ${\rm max} \left( m_{1,\chi}^2 \right) \sim {\rm max} \left( m_{2,\chi}^2 \right)$ but with the two terms out of phase with each other, so $\tilde{m}_{\rm eff,\chi}^2$ does not develop sharp features and the modes $z_k (t)$ do not undergo rapid amplification.

\begin{figure}
\centering
\includegraphics[width=.48\textwidth]{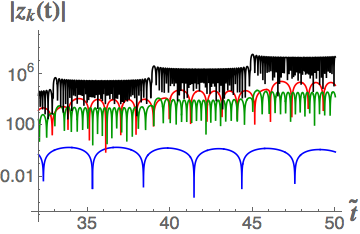}  \quad \includegraphics[width=.48\textwidth]{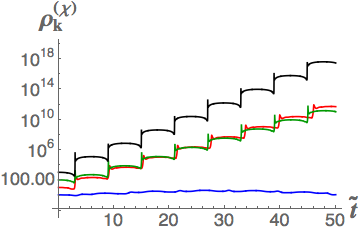} 
 \caption{\small \baselineskip 11pt
({\it Left}) The amplitude of the isocurvature mode $\vert z_k (t) \vert$ for $k = 0$, with $g / \lambda_\phi = 2$ and varying $\xi_\phi$. Note that as $\xi_\phi$ increases, $z_k (t)$ oscillates more frequently per oscillation of the background field, entering the broad-resonance regime. ({\it Right}) The isocurvature energy density $\rho^{(\chi)}_k$ for $k = 0$ and $g/\lambda_\phi=2$. For both plots, $\xi_{\phi}=10,10^2,10^3,10^4$  (blue, red, green, and black, respectively).  }
\label{fig:zrandomridge}
\end{figure}

Since we have rescaled time as $\tilde{t} = \sqrt{\lambda_\phi} \, M_{\rm pl} \, t / \xi_\phi$, the background field's oscillation period for $\xi_\phi \gg 1$ is given by $\tilde{T} = T / \xi_\phi \simeq 14.8$ (in units of $(\sqrt{\lambda_\phi} \, M_{\rm pl})^{-1}$). As shown in Fig.~\ref{fig:zrandomridge}, larger values for the isocurvature mass put the system into the broad-resonance regime \cite{KLS}, in the sense that the fluctuations oscillate multiple times for each background oscillation, with correspondingly rapid growth of $\rho^{(\chi)}_k$.

\subsubsection{Varying Ellipticity}

\begin{figure}
\centering
\includegraphics[width=0.48\textwidth]{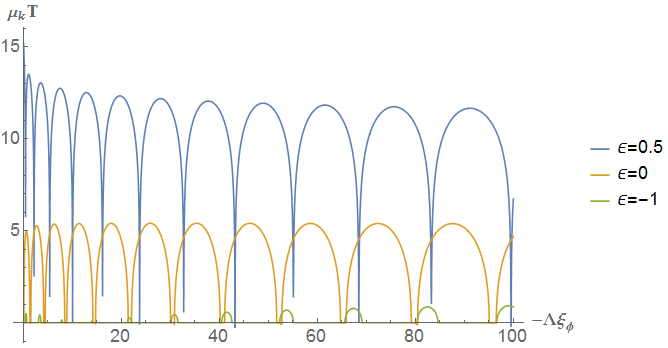}  \quad \includegraphics[width=0.48\textwidth]{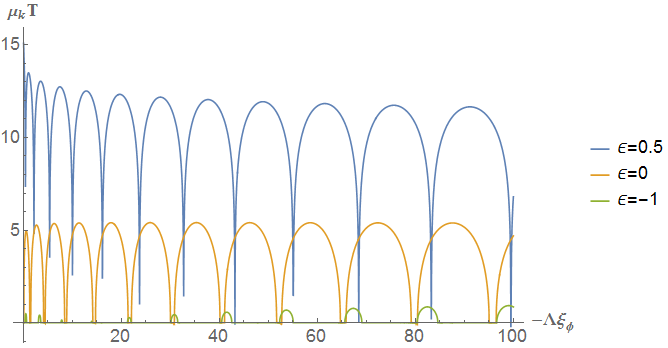}  
\\
\includegraphics[width=0.48\textwidth]{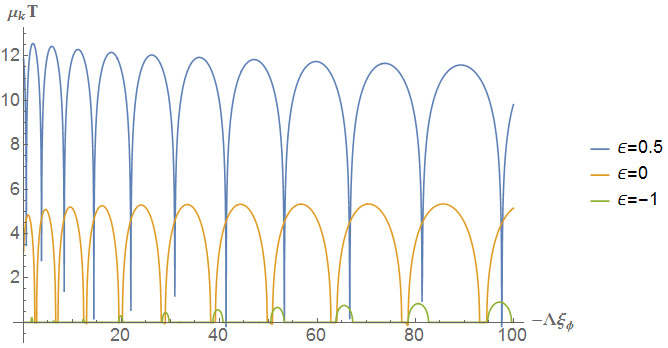} \quad  \includegraphics[width=0.48\textwidth]{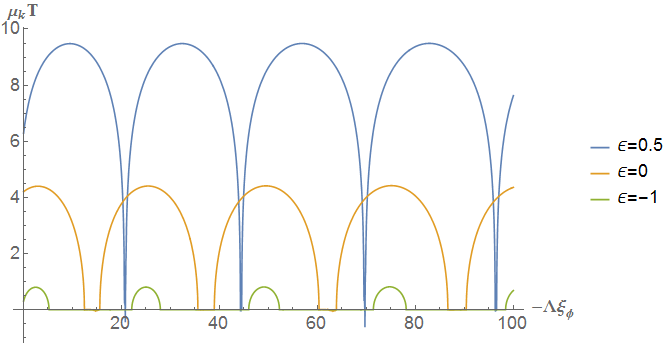}  
\caption{\small \baselineskip 11pt Floquet exponents for the isocurvature perturbations normalized by the period of background oscillations for $\xi_\phi = 10^4$ and varying $\varepsilon = 0.5,0,-1$. Each two-dimensional plot corresponds to a slice of the full Floquet chart at a specific normalized wavenumber: ({\it top, left to right}) $kT = 0,1$, ({\it bottom, left to right}) $kT = 10,100$. }
\label{fig:isocurv_ellipticity_large_xi}
\end{figure}

For completeness we construct the Floquet charts for elliptical potentials, with $\xi_\phi \ne \xi_\chi$. In Ref.~\cite{SSK} we introduced the ellipticity parameter $\varepsilon \equiv (\xi_\phi - \xi_\chi) / \xi_\phi$. Fig.\ \ref{fig:isocurv_ellipticity_large_xi} shows the resulting instability bands for varying ellipticity. We fix $\xi_\phi=10^4$, which is well into the regime of large nonminimal coupling. For positive ellipticity, meaning $\xi_\chi<\xi_\phi$, the instability bands become larger both in amplitude and width. The opposite occurs for negative ellipticity, with $\xi_\chi > \xi_\phi$, for which the instability bands become suppressed and further apart. For $\xi_\chi = 2\xi_\phi$ (or $\varepsilon = -1$), the resonances vanish for most values of the effective coupling $- \Lambda_\phi \xi_\phi$. This behavior during preheating is reminiscent of the behavior of these systems during inflation \cite{SSK}, for which the fraction of isocurvature modes produced, $\beta_{\rm iso}$, is enhanced for positive ellipticity and suppressed for negative ellipticity \cite{ellipticfn}. 

\section{Conclusions}
\label{Conclusions}

Using the covariant formalism developed in Ref.~\cite{MultiPreheat1}, in this paper we have investigated the resonance structure for the amplification of adiabatic and isocurvature perturbations during preheating in models with multiple scalar fields nonminimally coupled to gravity. In these models, the background dynamics generically fall into a single-field attractor that persists (at least) through the preheating phase \cite{MultiPreheat1}, thereby avoiding the ``de-phasing" \cite{Barnaby,Battefeld} that commonly occurs in multifield models with minimal couplings. Within the approximation of a rigid spacetime, we identified several unique features of the preheating dynamics that arise due to the nontrivial field-space manifold, which have no analogue in minimally coupled models.

For nonminimal couplings $\xi_I \geq {\cal O} (1)$, the spectral content of the background field differs significantly from its minimally coupled counterpart, the well-known quartic model. Whereas in the case of the minimally coupled quartic model the background dynamics may be well described by the first two harmonics, in the presence of nonminimal couplings higher harmonics attain comparable magnitude to the lowest harmonics. Furthermore, for a specific combination of the field's amplitude and nonminimal coupling, all but the fundamental harmonic vanish, giving a simple sinusoidal solution. For large nonminimal couplings, the ratios of the Fourier coefficients for various harmonics quickly asymptote to fixed values, such that for $\xi_\phi > {\cal O}(100)$ the spectral content of the background inflaton field is independent of the exact value of  $\xi_\phi$.
 
The behavior of the adiabatic and isocurvature perturbations is very different. For the adiabatic modes, the Floquet charts for small nonminimal couplings present a dominant, primary  instability band at nonzero values of the wavenumber. This is completely reversed for large nonminimal couplings, for which the primary instability band occurs for $k \sim 0$, making the two cases --- small and large nonminimal coupling --- qualitatively distinct. Furthermore, for large nonminimal coupling $\xi_\phi \gtrsim 10^3$, the Floquet chart quickly asymptotes to a common shape, yielding a single scaling behavior for the adiabatic perturbations in the limit of large $\xi_I$.

The isocurvature modes show richer phenomenology, since their behavior also depends on the coupling between the two fields $\phi$ and $\chi$. For small values of the nonminimal coupling the Floquet chart resembles the Lam\'{e} chart, arising in the study of a minimally coupled quartic model. As the value of the nonminimal coupling increases, the instability bands tilt and become narrower, thus making preheating less efficient for $0 < \xi_I < 1$ compared to the case with $\xi_I = 0$. For large values of the nonminimal coupling, on the other hand, the Floquet chart is comprised of a dense set of closely spaced, almost parallel instability bands, making the amplification of isocurvature modes very efficient. Furthermore, the scaling solution found in the adiabatic case is also present for isocurvature modes, although the single behavior in the $\xi_I \rightarrow \infty$ limit is reached more slowly with increasing $\xi_I$ than in the adiabatic case. 

Finally, the preheating dynamics for both adiabatic and isocurvature modes reveal an intermediate region, for $\xi_I \sim {\cal O} (1 - 10)$. Within this intermediate region, the resonance structure for both adiabatic and isocurvature modes shows sensitive dependence on wavenumber and couplings, distinct from the minimally coupled case while also quite different from the behavior for $\xi_I \geq 10^2$. The emergence of this intermediate region during preheating is distinct from the behavior of spectral observables during inflation, such as the spectral index ($n_s$) and tensor-to-scalar ratio ($r$), which attain their large-$\xi_I$ values for $\xi_I \gtrsim {\cal O} (10)$ \cite{KMS,GKS,KS,SSK}. Thus post-inflation dynamics might provide one means of breaking the observational degeneracy of this class of models, even given the strong single-field attractor behavior during and after inflation.

Our aim in the present paper has been to understand how the resonance structure for this class of models changes with the strength of the nonminimal couplings, within the rigid-spacetime approximation. We have therefore focused on the early preheating phase of reheating, working to linear order in the fluctuations and neglecting significant nonlinear effects such as backreaction from produced particles. We have found efficient resonance for $\xi_I \geq {\cal O} (100)$. In Ref.~\cite{MultiPreheat3} we relax the assumption of a rigid spacetime, and consider some possible observational consequences of such an efficient preheating phase.

\section*{Appendix A: Asymptotic equation of motion}
\label{AppendixAsymEOM}

Starting from Eq. \eqref{eq:asymptoticeq} and performing the folllowing change of variables,
\begin{align}
x = \log(1+\delta^2) ,
\end{align}
the term including the first derivative vanishes and the resulting equation is
\begin{align}
\ddot x + {1 \over 3} e^{-2x} (e^x-1)=0 \, ,
\label{xeom}
\end{align}
where we work in terms of $\tilde{t} = \sqrt{\lambda_\phi} \, M_{\rm pl} \, t / \xi_\phi$ and overdots denote $d / d\tilde{t}$. Eq.~(\ref{xeom}) is an analytically solvable asymptotic equation of motion which may be integrated in two steps. The first integration yields
\beq
 {\dot x d\dot x } =   {1 \over 3} e^{-2x} (1-e^x) \> dx \> ,
 \eeq
which integrates to
\beq
\dot x = \sqrt{ 2 \left ( C + \frac{1}{3} e^{-x} - \frac{1}{6} e^{-2x}  \right ) } \> .
\label{xint2}
\eeq 
The constant of integration $C$ is defined by setting $\dot x =0$ for $x=\log(1+\alpha^2)$, where $\alpha$ is defined via $\phi_{\rm max}  = \alpha M_{\rm pl} /\sqrt{\xi_\phi}$, where $\phi_{\rm max}$ is the maximum field amplitude during the oscillation phase. By matching to the inflationary solution near $\tilde{t}_{\rm end}$, $\alpha$ is constrained to $\alpha \le 0.8$. Eq.~(\ref{xint2}) may then be integrated:
\beq
\frac{e^x \sqrt{ 6 C + e^{-2x} (2 e^x - 1) } \> \ln [ 1 + 6 C e^x + \sqrt{ (6C) (6C e^{2x} + 2 e^x - 1)} ] }{\sqrt{ (2C)( 6 C e^{2x} + 2 e^x - 1) } } = \tilde{t} + {\cal C} .
\label{xsol}
\eeq
The new integration constant ${\cal C}$ may be determined by setting $x(0)=\log(1+\alpha^2)$. Finally we may revert to our original variable by using $x=\log(\delta^2+1)$ and $\delta = \sqrt{\xi_\phi} \> \phi / M_{\rm pl}$. 

This is anything but a simple formula, and it cannot be analytically inverted to give $\delta(\tilde{t})$. Nonetheless, we have demonstrated the existence of an oscillatory solution whose behavior is independent of $\xi_\phi$ in the limit $\xi_\phi \rightarrow \infty$. 
Hence we may solve the asymptotic equation of motion (with no ambiguity about the value of $\xi_\phi$) and calculate $\delta(\tilde{t})$ numerically, as in Section~\ref{BackgroundSpectral}. The spectrum of $\delta(\tilde{t})$ is easily derived and compared with the spectrum of $\phi(\tilde{t})$ for finite values of $\xi_\phi$, as in Fig.\ \ref{anfig}, showing excellent agreement for $\xi_\phi \gg 1$.

\section*{Appendix B: Semi-Analytic calculation of Floquet-band Boundaries}
\label{AppendixFloquet}

In this Appendix we construct infinite-dimensional matrices, whose vanishing determinants may be used to determine the stability boundaries, where ${\rm Re} [\mu_k] = 0$, for both adiabatic and isocurvature modes. We also demonstrate that the characteristics of the boundary for the primary instability band may be determined with sufficient accuracy by truncating the matrices to simpler $3 \times 3$ form.

\subsubsection{$T/2$-Periodic Solutions}

For convenience we define $\omega' \equiv 4 \pi / T = 2 \omega$, where $T$ is the period of the background field's oscillations. We may then make the following expansions for the mode functions and their effective frequencies:

\beqn
\begin{split}
v_k &= \displaystyle\sum_{j=0}^\infty \alpha_{k,j} \cos{2j \omega t}, \quad z_k = \displaystyle\sum_{j=0}^\infty \beta_{k,j} \cos{2j \omega t} \\
\Omega_{(\phi)}^2 &= k^2 + a_{k,0} + \displaystyle\sum_{j=1}^\infty a_{k,j} \cos{2j \omega t} ,\quad a_{k,j} = \left(1 - {\delta_{j,0} \over 2} \right) {4 \over T}\displaystyle\int_{0}^{T/2} m_{{\rm eff},\phi}^2 \cos{ 2 j \omega t} \\
\Omega_{(\chi)}^2 &= k^2 + b_{k,0} + \displaystyle\sum_{j=1}^\infty b_{k,j} \cos{2j \omega t} , \quad b_{k,j} = \left(1 - {\delta_{j,0} \over 2} \right) {4 \over T}\displaystyle\int_{0}^{T/2} m_{{\rm eff},\chi}^2 \cos{ 2 j \omega t} .
\end{split}
\eeqn
Plugging these relations into the equation of motion for the adiabatic modes, $v_k$, in Eq.~(\ref{vzeom}) (and taking the rigid-spacetime limit, so that $\eta \rightarrow t$), we find 
\begin{align}
\nonumber 0 &= \displaystyle\sum_{m=0}^\infty (-4m^2 \omega^2 + k^2 + a_{k,0}) \alpha_{k,m} \cos{2 m \omega t} + \displaystyle\sum_{m, n=1}^\infty a_{k,m}\alpha_{k,n} \cos{2 m \omega t} \cos{2 n \omega t}  \\
\nonumber &= \displaystyle\sum_{m=0}^\infty (-4m^2 \omega^2 + k^2 + a_{k,0}) \alpha_{k,m} \cos{2 m \omega t} + \displaystyle\sum_{m, n=1}^\infty {a_{k,m}\alpha_{k,n} \over 2} [\cos{2 (m + n) \omega t} + \cos{2(m-n) \omega t} ]  \\ 
&= \displaystyle\sum_{p=0}^\infty \gamma_p \cos{2p\omega t} ,
\end{align}
where the $\gamma_p$ are linear combinations of the $\alpha_{k,m}$ coefficients. The boundaries of the stability regions for the modes $v_k$ correspond to those places where each $\gamma_p = 0$. These correspond to the row of a matrix, $U$, whose vanishing determinant enforces the equation of motion of Eq.~(\ref{vzeom}). The elements of the matrix $U$ are given by 
\beqn
\begin{split}
U_{00} &=  k^2 + a_{k,0} \\
U_{p,p} &= -4p^2\omega^2 + k^2 + a_{k,0} + {a_{k,2p} \over 2}, \quad p\ge 1 \\
U_{p,q} &= {a_{k,|p-q|} + a_{k,p+q} \over 2}, \quad p\ne q
\end{split}
\eeqn
The corresponding matrix for the isocurvature modes, $z_k$, which we will call $Z$, is given by simply replacing each $a$ in the matrix above with $b$:
\beqn
\begin{split}
Z_{00} &=  k^2 + b_{k,0} \\
Z_{p,p} &= -4p^2\omega^2 + k^2 + b_{k,0} + {b_{k,2p} \over 2}, \quad p\ge 1 \\
Z_{p,q} &= {b_{k,|p-q|} + b_{k,p+q} \over 2}, \quad p\ne q
\end{split}
\eeqn

We may also expand $v_k$ and $z_k$ into their respective sine series:
\begin{align}
v_k &= \displaystyle\sum_{j=1}^\infty \alpha_{k,j} \sin{2j \omega t}, \quad z_k = \displaystyle\sum_{j=1}^\infty \beta_{k,j} \sin{2j \omega t} 
\end{align}
With these expansions, Eq.~\eqref{vzeom} becomes 
\begin{align}
\nonumber
0 &= \displaystyle\sum_{m=1}^\infty (-4m^2 \omega^2 + k^2 + a_{k,0}) \alpha_{k,m} \sin{2 m \omega t} + \displaystyle\sum_{m, n=1}^\infty a_{k,m}\alpha_{k,n} \cos{2 m \omega t} \sin{2 n \omega t}  \\
\nonumber
&= \displaystyle\sum_{m=1}^\infty (-4m^2 \omega^2 + k^2 + a_{k,0}) \alpha_{k,m} \sin{2 m \omega t} + \displaystyle\sum_{m, n=0}^\infty {a_{k,m}\alpha_{k,n} \over 2} [\sin{2 (n+m) \omega t} + \sin{2(n-m) \omega t} ]  \\ 
&= \displaystyle\sum_{p=1}^\infty \zeta_p \sin{2p\omega t} .
\end{align}
Just as for the cosine expansion, we find a system of equations by setting each $\zeta_p = 0$. The matrix $\mathcal{U}$ is given by
\beqn
\begin{split}
\mathcal{U}_{p,p} &= -4p^2\omega^2 + k^2 + a_{k,0} - {a_{k,2p} \over 2}, \quad p\ge 1 \\
\mathcal{U}_{p,q} &= {a_{k,{|p-q|}} + \text{sign}(q-p) a_{k,p+q} \over 2}
\end{split}
\eeqn
The corresponding matrix for the $z_k$'s is represented by the matrix $\mathcal{Z}$. The form of $\mathcal{Z}$ is exactly the same as that of $\mathcal{U}$ under the replacement of $a_{k,j }$ with $b_{k,j}$ for all $j$. 
 
\subsubsection{$T$-Periodic Solutions}

We now consider the case in which $v_k$ and $z_k$ are $T$-periodic.  We can then make the following expansions:
\begin{align}
v_k &= \displaystyle\sum_{j=0}^\infty \alpha_{k,j} \cos{j \omega t}, \quad z_k = \displaystyle\sum_{j=0}^\infty \beta_{k,j} \cos{j \omega t} .
\end{align}
Plugging these relations into Eq. \eqref{vzeom}, we find 
\begin{align}
\nonumber
0 &= \displaystyle\sum_{m=0}^\infty (-m^2 \omega^2 + k^2 + a_{k,0}) \alpha_{k,m} \cos{ m \omega t} + \displaystyle\sum_{m, n=1}^\infty a_{k,m}\alpha_{k,n} \cos{2 m \omega t} \cos{n \omega t}  \\
\nonumber
&= \displaystyle\sum_{m=0}^\infty (-m^2 \omega^2 + k^2 + a_{k,0}) \alpha_{k,m} \cos{m \omega t} + \displaystyle\sum_{m, n=1}^\infty {a_{k,m}\alpha_{k,n} \over 2} [\cos{(2m + n) \omega t} + \cos{(2m-n) \omega t} ]  \\ 
&= \displaystyle\sum_{p=0}^\infty \delta_p \cos{2p\omega t}  \, ,
\end{align}
where $\delta_p$ are linear combinations of the $\alpha_{k,m}$ coefficients. Just as in the previous case, we form a matrix $U'$ whose vanishing determinant yields equations for the boundaries of the stability bands for the adiabatic modes, $v_k$: 
\begin{align}
U'_{00} &=  k^2 + a_{k,0} \\
U'_{p,p} &= -p^2\omega^2 + k^2 + a_{k,0} + {a_{k,p} \over 2}, \quad p\ge 1 \\
U'_{p,q} &= {a_{k,|p-q|/2} + a_{k,(p+q)/2} \over 2}, \quad p\ne q, p\equiv q\mod{2}
\end{align}
The corresponding matrix for $z_k$, which we call $Z'$, may be constructed by replacing $a_{k,j}$ with $b_{k,j}$ for all $ j $.

We must also consider sine series with period $T$. After a similar procedure, we find the matrices of interest:
\beqn
\begin{split}
\mathcal{U}'_{p,p} &= -p^2\omega^2 + k^2 + a_{k,0} - {a_{k,p} \over 2}, \quad p\ge 1 \\
\mathcal{U}'_{p,q} &= {a_{k,{|p-q|/2}} + \text{sign}(q-p) a_{k,(p+q)/2} \over 2}
\end{split}
\eeqn
and the same for $\mathcal{Z}'$ with $a_{k,j} \rightarrow b_{k,j}$. 

\begin{figure}
\centering
\includegraphics[width=0.48\textwidth]{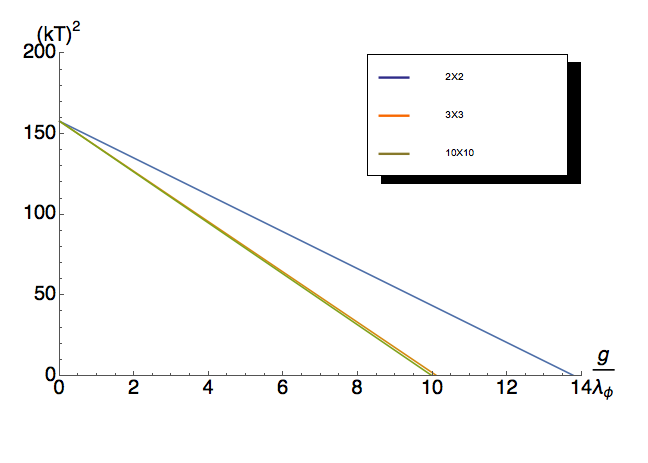} \quad \includegraphics[width=0.48\textwidth]{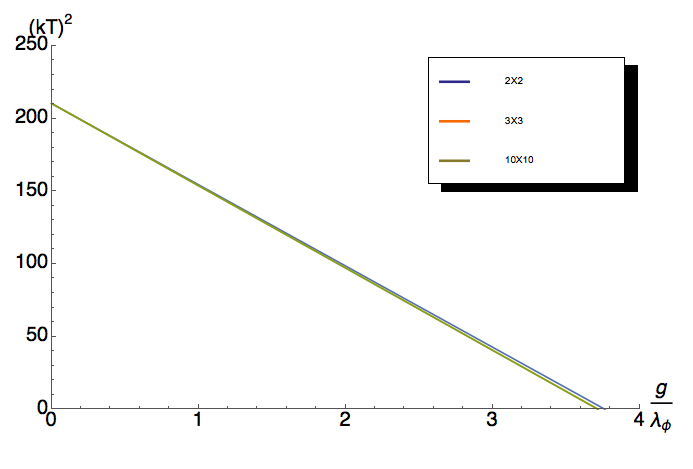}
\caption{\small \baselineskip 11pt Comparison of the stability boundary of the primary instability band given by $\det U = 0$ when $U$ is truncated to $2 \times 2$, $3 \times 3$, and $10 \times 10$ matrices, for $\xi_\phi = 0$ ({\it left}) and $\xi_\phi = 0.5$ ({\it right}). All higher-dimension truncations are too close to the $10 \times 10$ truncation to resolve. Hence we find that the $3 \times 3$ truncation is sufficient to yield an accurate approximation to the location of the stability boundary in the $(kT)^2 - (g / \lambda_\phi)$ plane. }
\label{fig:matrixdim}
\end{figure}

To summarize, we construct eight distinct matrices: four $U$-matrices for the adiabatic modes $v_k$ (coming from the $T/2$-periodic sine and cosine series and the $T$-periodic sine and cosine series), and four $Z$-matrices for the isocurvature modes $z_k$. Each of these matrices is a function of $g / \lambda_\phi$ and $k$. The vanishing of the determinants of these matrices determines the boundaries of the stability regions for the $v_k$ and $z_k$ modes in the $(kT)^2 - (g / \lambda_\phi)$ plane. Finally, in Fig.~\ref{fig:matrixdim}, we demonstrate that results of sufficient accuracy for the primary instability band may be calculated by truncating the infinite-dimensional matrices to simple $3 \times 3$ matrices. 

\acknowledgements{It is a pleasure to thank Mustafa Amin, Bruce Bassett, Jolyon Bloomfield, Peter Fisher, Tom Giblin, Alan Guth, Mark Hertzberg, Johanna Karouby, and an anonymous referee for helpful discussions. We would like to acknowledge support from the Center for Theoretical Physics at MIT. This work is supported by the U.S. Department of Energy under grant Contract Number DE-SC0012567. MPD and AP were also supported in part by MIT's Undergraduate Research Opportunities Program (UROP).  
CPW thanks the University of Washington College of Arts \& Sciences for financial support. She also gratefully acknowledges support from the MIT Dr. Martin Luther King, Jr. Visiting Professors and Scholars program and its director Edmund Bertschinger. EIS gratefully acknowledges support from a Fortner Fellowship at the University of Illinois at Urbana-Champaign. }


\begin{thebibliography}{999}
\bibitem{ellipticfn} Note that the definition of the ellipticity is defined globally for the system, regardless of the fields' trajectory through field space. This means that for a field trajectory along $\phi=0$, the effects for positive and negative ellipticity will be inverted, compared to Fig.\ \ref{fig:isocurv_ellipticity_large_xi}.
\end{thebibliography}
\end{document}